\documentclass[onecolumn,epjc3]{svjour3}  
\smartqed  
\RequirePackage{graphicx}

\RequirePackage{subfig}
\usepackage[a4paper,top=3cm,bottom=2cm,left=2cm,right=2cm,marginparwidth=1.75cm]{geometry}
\journalname{Eur. Phys. J. C}
\usepackage{lineno}

\begin{document}

\title{Performance and Moli\`ere radius measurements using a compact prototype of LumiCal in an electron test beam
}


\author{H.~Abramowicz\thanksref{addr1}
        \and
       A.~Abusleme\thanksref{addr2} 
 \and
     K.~Afanaciev\thanksref{addr3}
 \and
    Y.~Benhammou\thanksref{addr1}
 \and
     O.~Borysov\thanksref{addr1}
 \and
     M. Borysova\thanksref{e2,addr1}
 \and
      I.~Bozovic-Jelisavcic\thanksref{addr4}
 \and
     W.~Daniluk\thanksref{addr5}
 \and
     D.~Dannheim\thanksref{addr6}
 \and
     M.~Demichev\thanksref{addr7}
     \and
    K.~Elsener\thanksref{addr6}
 \and
    M.~Firlej\thanksref{addr8}
 \and
    E.~Firu\thanksref{addr9}
 \and
     T.~Fiutowski\thanksref{addr8}
 \and
     V.~Ghenescu\thanksref{addr9}
 \and
     M.~Gostkin\thanksref{addr7}
 \and
    M.~Hempel\thanksref{e1,addr10}
 \and
     H.~Henschel\thanksref{addr10}
 \and
    M.~Idzik\thanksref{addr8}
 \and
    A.~Ignatenko\thanksref{e3,addr3}
 \and
    A.~ Ishikawa\thanksref{addr11}
 \and
      A.~Joffe\thanksref{addr1}
      \and
      G.~Kacarevic\thanksref{addr4}
 \and
      S.~Kananov\thanksref{addr1}
 \and
     O.~Karacheban\thanksref{e1,addr10}
\and
    W.~Klempt\thanksref{addr6}
\and
 S.~Kotov\thanksref{addr7}
\and
    J.~Kotula\thanksref{addr5}
\and
   U.~Kruchonak\thanksref{addr7}
\and
    Sz.~Kulis\thanksref{addr6}
\and
   W.~Lange\thanksref{addr10}
\and
    J.~Leonard\thanksref{addr10}
\and
    T.~Lesiak\thanksref{addr5}
\and
   A.~Levy\thanksref{addr1}
\and
     I.~Levy\thanksref{addr1}
\and
   L.~Linssen\thanksref{addr6}
   \and
   W.~Lohmann\thanksref{e1,addr10}
\and
    J.~Moron\thanksref{addr8}
\and
    A.~Moszczynski\thanksref{addr5}
\and
    A.T.~Neagu\thanksref{addr9}
  \and
    B.~Pawlik\thanksref{addr5}
    \and
   T.~Preda\thanksref{addr9}
     \and
   A.~Sailer\thanksref{addr6}
\and
    B.~Schumm\thanksref{addr12}
 \and
    S.~Schuwalow\thanksref{e4,addr10}
   \and
    E.~Sicking\thanksref{addr6}
    \and
    K.~Swientek\thanksref{addr8}
   \and
        B.~Turbiarz\thanksref{addr5}
        \and
        N.~Vukasinovic\thanksref{addr4}
        \and
    T.~Wojton\thanksref{addr5}
   \and
    H.~Yamamoto\thanksref{addr11}
   \and
     L.~Zawiejski\thanksref{addr5}
   \and
      I.S.~Zgura\thanksref{addr9}
   \and
    A.~Zhemchugov\thanksref{addr7}
}

\thankstext{e2}{Visitor from Institute for Nuclear Research NANU (KINR), Kyiv 03680, Ukraine.}
\thankstext{e1}{Also at Brandenburg University of Technology, Cottbus, Germany.}
\thankstext{e3}{Now at DESY, Zeuthen, Germany.}
\thankstext{e4}{Also at DESY, Hamburg, Germany.}


\institute{
      Raymond \& Beverly Sackler School of Physics \& Astronomy, Tel Aviv University, Tel Aviv, Israel\label{addr1}
 \and
      Pontificia Universidad Catolica de Chile, Santiago, Chile \label{addr2}
 \and
      NC PHEP, Belarusian State University, Minsk, Belarus \label{addr3}
      \and
      Vinca Institute of Nuclear Sciences, University of Belgrade, Belgrade, Serbia \label{addr4}
      \and
  IFJ PAN, PL-31342, Krakow, Poland \label{addr5}
 \and
    CERN, Geneva, Switzerland\label{addr6}
    \and
     JINR, Dubna, Russia \label{addr7}
 \and
   Faculty of Physics and Applied Computer Science, AGH University of Science and Technology, Krakow, Poland \label{addr8}
 \and
     ISS, Bucharest, Romania \label{addr9}
     \and
      DESY, Zeuthen, Germany \label{addr10}
 \and
   Tohoku University, Sendai, Japan \label{addr11}
 \and
     University of California, Santa Cruz, USA\label{addr12}
}



\maketitle

\abstract{A new design of a detector plane of sub-millimetre thickness for an electromagnetic sampling calorimeter is presented. It is intended to be used in the luminometers LumiCal and BeamCal in future linear e$^{+}$e$^{-}$ collider experiments. The detector planes were produced utilising novel connectivity scheme technologies. They were installed in a compact prototype of the calorimeter and tested at DESY with an electron beam of  energy 1-5~GeV. The performance of a prototype of a compact LumiCal comprising eight detector planes was studied. The effective Moli\`ere radius at 5~GeV was determined to be (8.1 $\pm$ 0.1 (stat) $\pm$ 0.3 (syst))~mm, a value well reproduced by the Monte Carlo (MC) simulation (8.4 $\pm$ 0.1)~mm. The
dependence of the effective Moli\`ere radius on the electron energy in the range 1-5~GeV was also studied. Good agreement was obtained between data and MC simulation.}





\flushbottom

\section{Introduction}
\label{Introduction}
Forward calorimeters for future electron positron linear collider experiments have challenging requirements on a fast and high precision measurement of the luminosity~\cite{FCAL_ILC}, resulting in a stringent set of specifications for highly compact calorimeters. Two such calorimeters, LumiCal and BeamCal, are being considered for installation in the forward region of both International Linear Collider(ILC)~\cite{ILC_TDR_v4_det,ILC_TDR_v1_phys} detectors, ILD and SiD, and also in the Compact Linear Collider (CLIC) detector~\cite{CLIC_UPDATE_YP}. 
The precise measurement of the integrated luminosity is provided by the LumiCal detector. BeamCal is designed for instant luminosity measurement and beam-tuning when included in a fast feedback system as well as for tagging beam particles scattered through low angles. Both detectors extend the capabilities of the experiments for physics studies in the high rapidity region. 
\begin{figure}[h!]
  \begin{minipage}[c]{0.50\textwidth}
    \includegraphics[width=\textwidth]{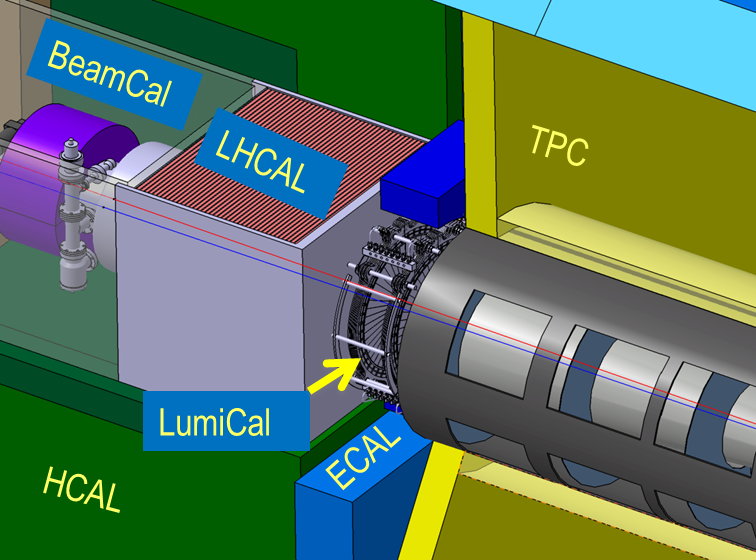}
  \end{minipage}\hfill
  \begin{minipage}[t]{0.45\textwidth}
    \caption{The very forward region of the ILD detector. LumiCal, BeamCal and LHCAL are carried by the support tube 
             for the final focusing quadrupole and the beam-pipe. TPC, ECAL and HCAL are the Time Projection Chamber and the Electromagnetic and Hadron Calorimeters.
    } \label{fig_forward_ild}
  \end{minipage}
\end{figure}

The layout of one arm of the forward region of the ILD detector is presented in Figure~\ref{fig_forward_ild}. LumiCal is positioned in a circular hole of the end-cap electromagnetic calorimeter ECAL. BeamCal is placed just in front of the final focus quadrupole. LumiCal is designed as a sampling calorimeter composed of 30 layers of 3.5~mm (1X$_0$) thick tungsten absorbers and silicon sensors placed in 
a one-millimeter gap between absorber plates. BeamCal has a similar design as LumiCal. For the current BeamCal baseline design, GaAs sensors are considered which can withstand  higher radiation doses at room temperature. The similarity between LumiCal and BeamCal designs implies that the technology developed for one can be used also for the other. 

Luminosity in LumiCal is measured using Bhabha scattering, e$^{+}$e$^{-}\rightarrow$~e$^{+}$e$^{-}$($\gamma$), as a gauge process. The Bhabha scattering cross section can be precisely calculated in QED~\cite{Bhabha_scatt} and the luminosity, $\sf{L}$, is obtained as 
\begin{equation}{
  {\sf{L}} = \frac{{N}_{\mathrm{B}}}{\sigma_{\mathrm{B}}},
  }\label{lumiDefEQ} 
\end{equation}
where $N_{\mathrm{B}}$ is the number of Bhabha events registered by LumiCal in a given range of polar angles~($\theta_{min}$,~$\theta_{max}$) and $\sigma_{\mathrm{B}}$ is the integral of the differential cross section over the same range. This range defines the fiducial volume of the calorimeter.  The fiducial volume for the LumiCal baseline design was studied in simulations~\cite{FCAL_ILC} and found to be in the range from 41 to 67~mrad while the geometrical coverage of the LumiCal ranges from 31 to 77~mrad. The fiducial volume is reduced due to the lateral energy leakage which depends on the electromagnetic shower development in the transverse plane. The compact design of the LumiCal with small gaps between absorber plates allows the transverse size of the shower to be kept small and to achieve in a relatively small $\theta$ angle range a sufficiently large fiducial volume for a precise luminosity measurement. It also improves the efficiency to detect electrons on top of a widely-spread background originating from beamstrahlung and two-photon processes.

In addition, the compact construction of LumiCal and BeamCal are essential to match the strict geometrical constraints imposed by the design of the detectors and accelerator needs near the interaction point. 

In an earlier test beam of a four-layer silicon-tungsten prototype of the LumiCal, an effective Moli\`ere radius\footnote{As we do not have a fully contained shower in the prototype of LumiCal used in the earlier and also this test beam, we measure an effective Moli\`ere radius.} of 24.0 $\pm$ 0.6~mm was measured~\cite{LumiCal_multilayer_tb2014_epjc}. The reason for this large value was a large air gap between the silicon sensor plane and the absorber plates because space was needed for a 3.5~mm thick readout board. 

In order to get a smaller Moli\`ere radius, it was essential to design, build and use  planes of sub-millimetre thickness to be inserted in a mechanical frame~\cite{mech_frame} in one millimetre gaps between the tungsten absorber plates.

This paper describes the design and construction of a compact LumiCal prototype calorimeter, hereafter referred to as calorimeter, and the results from test-beam measurements carried out at DESY, using an electron beam between~1~-~5~GeV energy. For the readout electronics, APV25 front-end boards~\cite{APV_ieee,APV_nima,APV_Hybrid} were used. 
The effective Moli\`ere radius of this compact configuration was calculated in a similar way to that in Ref.~\cite{LumiCal_multilayer_tb2014_epjc}. The energy dependence of the effective Moli\`ere radius in the energy range of~1~-~5~GeV is also measured. In addition, two sensor planes were put in front of the calorimeter to serve as tracker planes to distinguish between electrons and photons. The results of this latter study will be presented elsewhere.

\section{Thin detector plane construction}
\label{LumiCal_Thin_subsection}
\begin{figure}[h!]
  \begin{minipage}[t]{0.38\textwidth}
    \includegraphics[width=\textwidth]{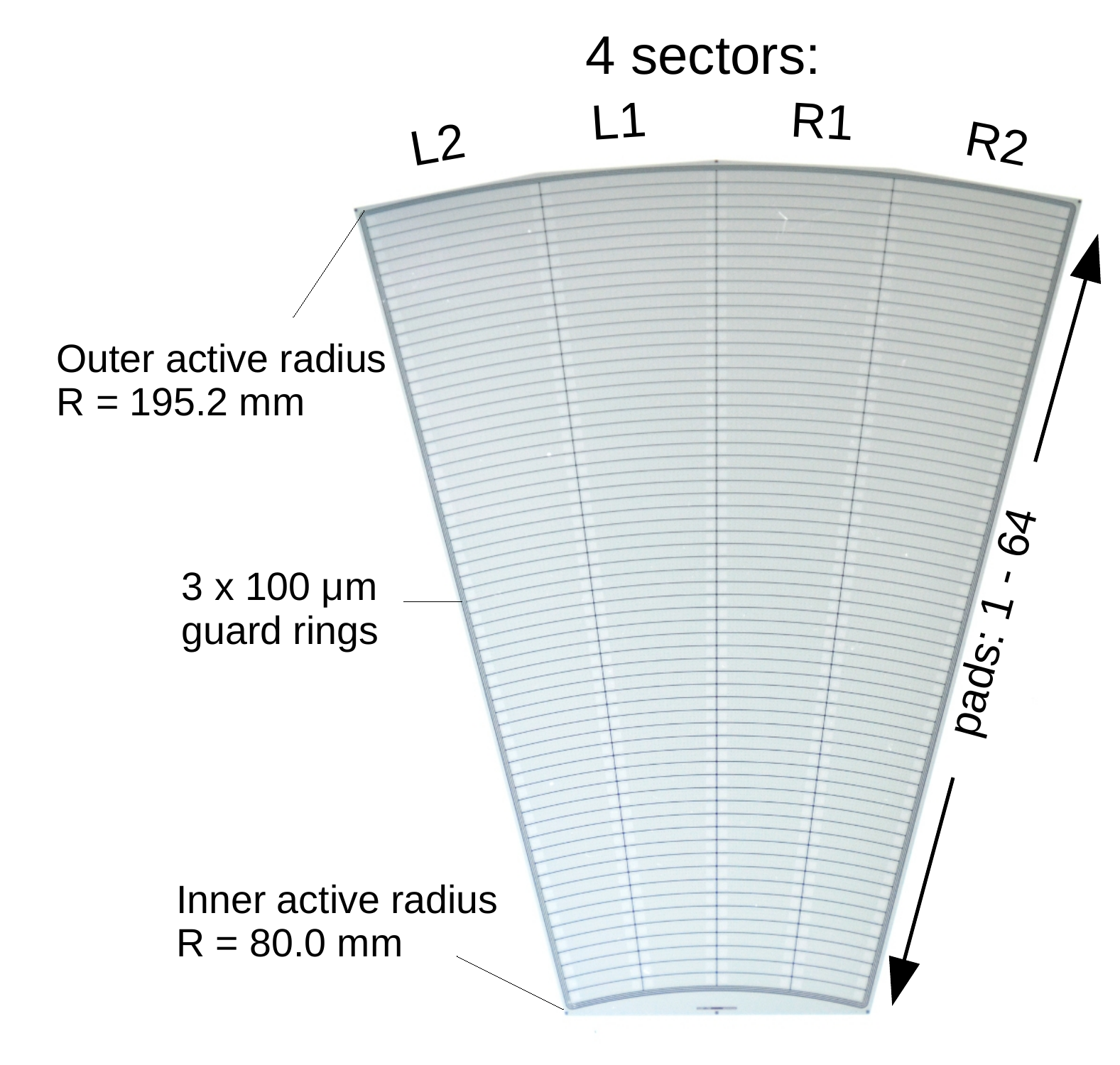}
    \caption{A LumiCal silicon sensor.}
    \label{fig_si_sensor}
  \end{minipage}\hfill
  \hspace{0.09\textwidth}
  \begin{minipage}[t]{0.52\textwidth}
    \includegraphics[width=0.99\columnwidth]{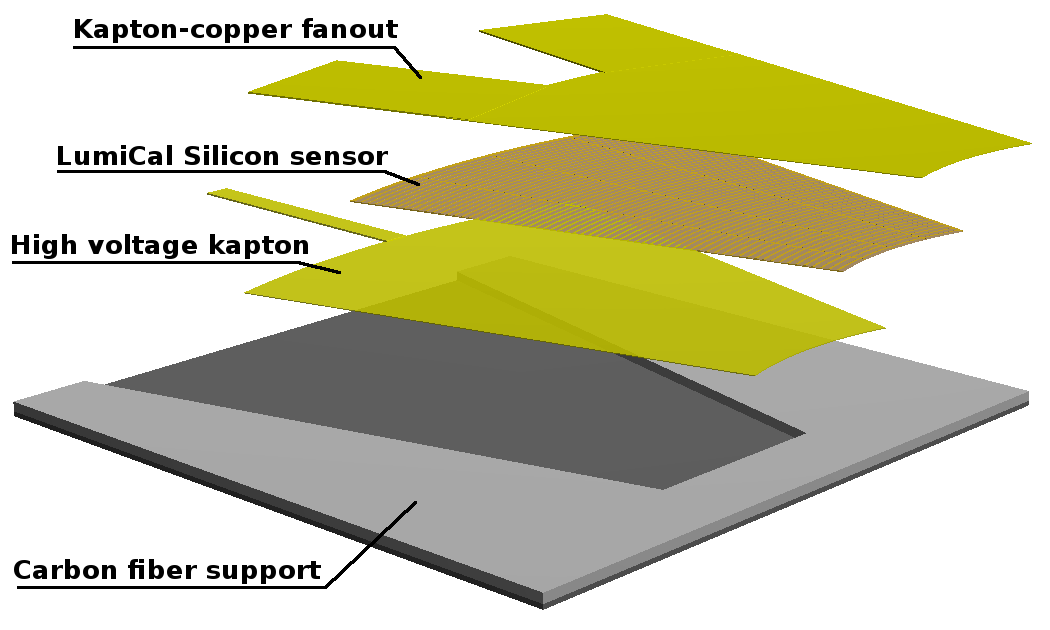}
    \caption{Detector plane assembly. The thickness of adhesive layers (not shown) between components is within 10~-~15~$\mu$m. The total thickness is 650~$\mu$m.}
    \label{fig_ThinLCAssembly}
  \end{minipage}\hfill
\end{figure}

The design of a LumiCal sensor was optimised in simulations to provide the required resolution of the polar angle reconstruction. A picture of a sensor is shown in Figure~\ref{fig_si_sensor}. The sensor is made of a 320~$\mu$m thick high resistivity n-type silicon wafer. It has the shape of a sector of a 30$^{\circ}$ angle, with inner and outer radii of the sensitive area of 80~mm and 195.2~mm, respectively. It comprises four sectors with 64 p-type pads of 1.8~mm pitch.  

The properties of the sensor were studied in the lab and beam tests. Results of beam tests and more details about the sensor can be found in Ref.~\cite{TB2010_jinst,LumiCal_multilayer_tb2014_epjc}. The first prototype of a LumiCal detector plane, which has been successfully used in a multi-layer configuration~\cite{LumiCal_multilayer_tb2014_epjc}, had a thickness of about 4~mm and only 32 pads were connected to the readout electronics. 

\begin{figure}[h!]
\centering
  \includegraphics[width=0.5\textwidth]{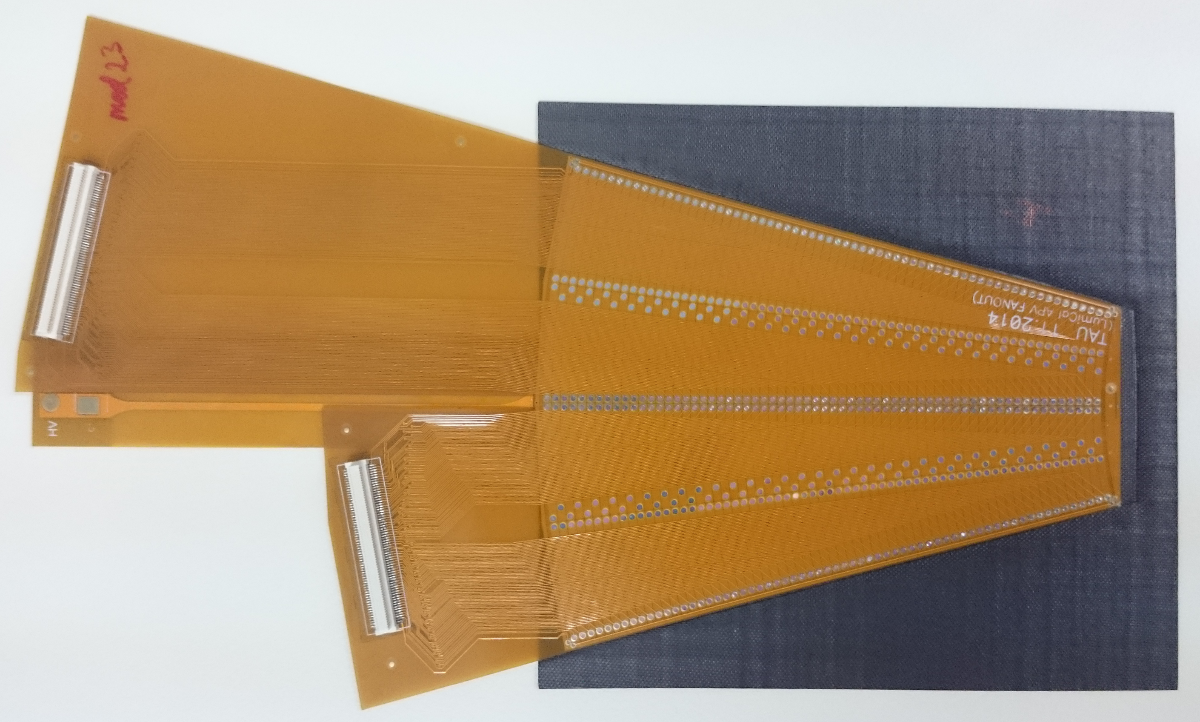}
    \caption{A thin detector plane. The black part is the carbon fibre support. The silicon sensor is covered by the Kapton fan-out which has two connectors for front-end boards.}
    \label{module_photo}
  \end{figure}

For the construction of a sub-millimetre detector plane we used the same silicon sensor.
The bias voltage is supplied to the n-side of the sensor 
by a 70~$\mu$m flexible Kapton-copper foil, glued to the sensor with a conductive glue.
The 256~pads of the sensor are connected to the front-end electronics using a fan-out made of 120~$\mu$m thick 
flexible Kapton foil with copper traces. The inner guard ring is grounded. Ultrasonic wire bonding was used to connect conductive traces on the 
fan-out to the sensor pads. A support structure, made of carbon fibre composite with a thickness of 100~$\mu$m in the sensor-gluing area, provides mechanical stability for the detector plane. Special fixtures were designed and produced to ensure the necessary thickness and uniformity of three glue layers between different components of the detector plane all over the area of the sensor. A sketch of the structure of the detector plane is shown 
in Figure~\ref{fig_ThinLCAssembly} and a photo of a completed plane in Figure~\ref{module_photo}. 
Since the multi-channel version of the dedicated front-end electronics is still under development, 
the APV25 front-end board~\cite{APV_ieee,APV_nima}, used by the silicon strip detector of the CMS experiment, was chosen as a temporary solution.
It has 128 channels, hence two boards read the whole sensor.  

The ultrasonic wire bonding proved to provide good electrical performance, but for a detector plane thinner than 1~mm, the wire loops, which are typically 100$-$200~$\mu$m high, cause a serious problem when the  plane needs to be installed in a 1~mm gap between absorber plates. The parameters of the bonding machine were studied and tuned to make the loop as low as possible and technically acceptable. The sampling based measurements, which were done using a con-focal laser scanning microscope, show that the loop height is in the range from 50~$\mu$m to~100~$\mu$m.

%
\section{Beam Test Setup}
\label{TB_setup}
\begin{figure}[h!]
  \centering
  \includegraphics[width=0.9\textwidth]{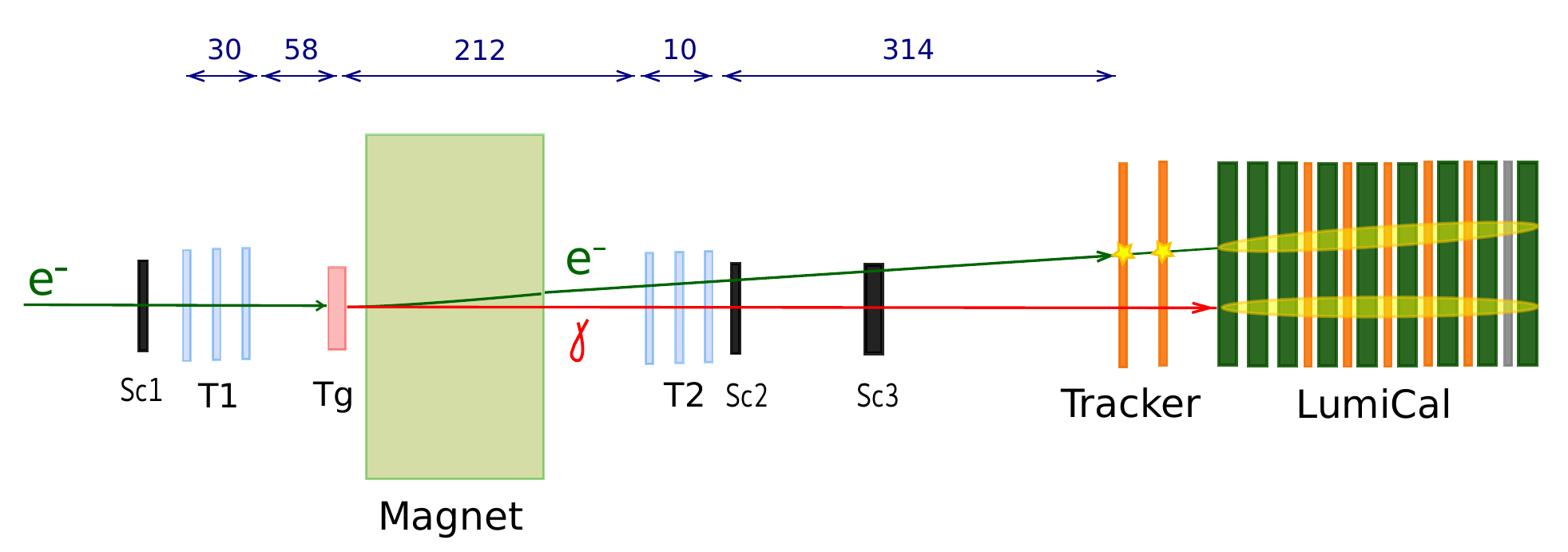}
  \caption{Geometry of the beam test setup (not to scale). Sc1, Sc2 and Sc3 are scintillator counters; 
           T1 and T2~ the arms of three-pixel detector planes, Tg~the copper target for bremsstrahlung photon production and LumiCal, the calorimeter prototype under test. Distances, rounded to integer numbers in centimetres, are shown in the upper part of the figure.}
  \label{tb2016}
\end{figure}

The detector planes were installed in the 1-mm gap between the tungsten absorber layers.
Each tungsten absorber layer is on average 3.5 mm thick and roughly one radiation length ($1~X_0$).
As described in Figure~\ref{tb2016}, the first calorimeter sensor layer was placed after 3 absorber layers, and the rest followed after each additional absorber layer. 
The last sensor layer was placed after 8 absorber layers with a total thickness of $7.7~X_0$, since, as noted in~\cite{LumiCal_multilayer_tb2014_epjc}, the absorber layers are not pure tungsten.
The detector planes were tested in two beam test campaigns in 2015 and 2016 at the DESY-II Synchrotron using electrons with energies between 1~GeV and 5~GeV. 

The beam test aimed to study the performance of the compact calorimeter   and to test the concept of tracking detectors in front of the calorimeter as a tool for electron and photon identification. The geometry of the setup is shown in Figure~\ref{tb2016}. 
The electron beam passed through a $5\times 5$ mm$^2$ square collimator that limits the beam spread along the test setup.
The AIDA/EUDET beam telescope was placed upstream of the calorimeter. 
The telescope was split into two parts T1 and T2, each containing an arm with 3 layers of MIMOSA-26 pixel silicon detectors and 2 thin 
scintillator counters Sc1 and Sc2, for the trigger system. The telescope front arm was placed before the dipole magnet to record the incoming electrons. 
The rear arm was placed after the dipole magnet to record the electrons in the direction of the calorimeter, and to separate them from the photons generated in the copper target that was mounted just in front of the magnet. 

\begin{figure}[h!]
 \centering
    \includegraphics[width=0.5\textwidth]{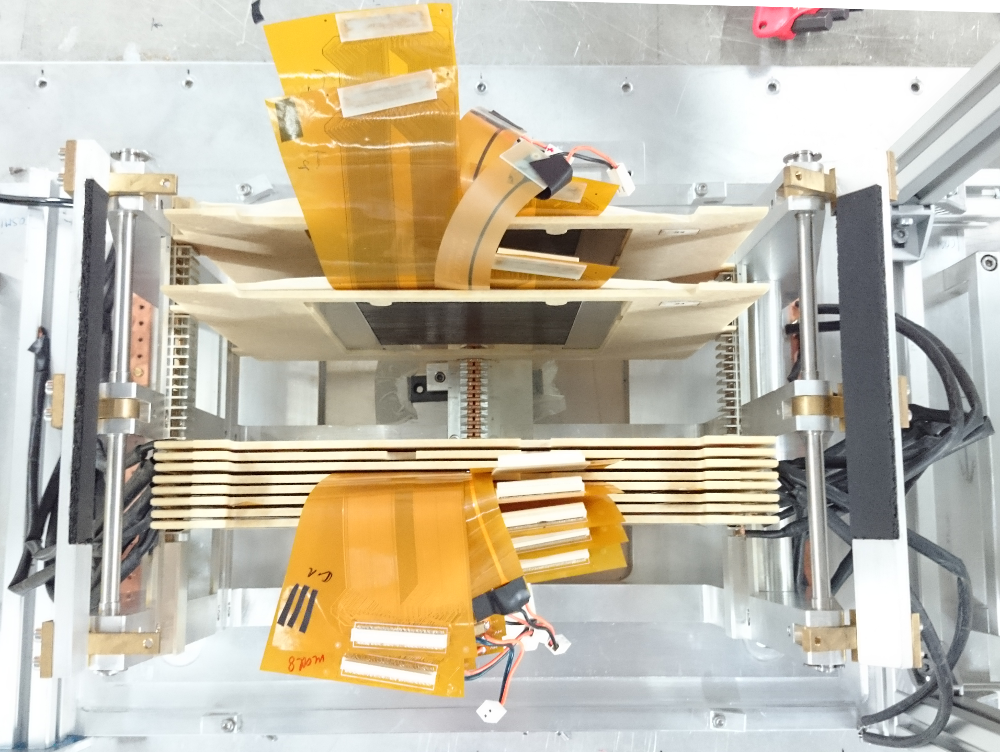}
    \caption{Top view of the assembled calorimeter.}
    \label{tb2016_LumiCal}
\end{figure}

The calorimeter  and tracker were assembled in a mechanical frame~\cite{mech_frame} specially designed to provide high precision positioning of the sensor planes and absorber plates. The sensor planes are attached to the tungsten absorber plates by adhesive tape. The tungsten plates are glued to permaglass inserted into the comb slots of the mechanical structure. The assembly of the calorimeter   is illustrated in Figure~\ref{tb2016_LumiCal}. Two sub-millimetre planes, viewed separately in the upper part of Figure~\ref{tb2016_LumiCal}, are the tracker planes denoted as ``Tracker'' in Figure~\ref{tb2016}. They are installed in front of the calorimeter. 

The last module in the LumiCal stack shown in grey is assembled using the tape automatic bonding (TAB) technology~\cite{TAB_Alice}. This sensor plane was not used in the present analysis.
All detector planes, for both the calorimeter  and the tracker, were powered with a reverse bias voltage of $120~V$.
This bias voltage is about 2 to 3 times the depletion voltage~\cite{TB2010_jinst}, but well below the breaking voltage of these silicon sensors. 
\section{Data Acquisition}

\begin{figure}[h!]
  \center
  \includegraphics[width=0.8\textwidth]{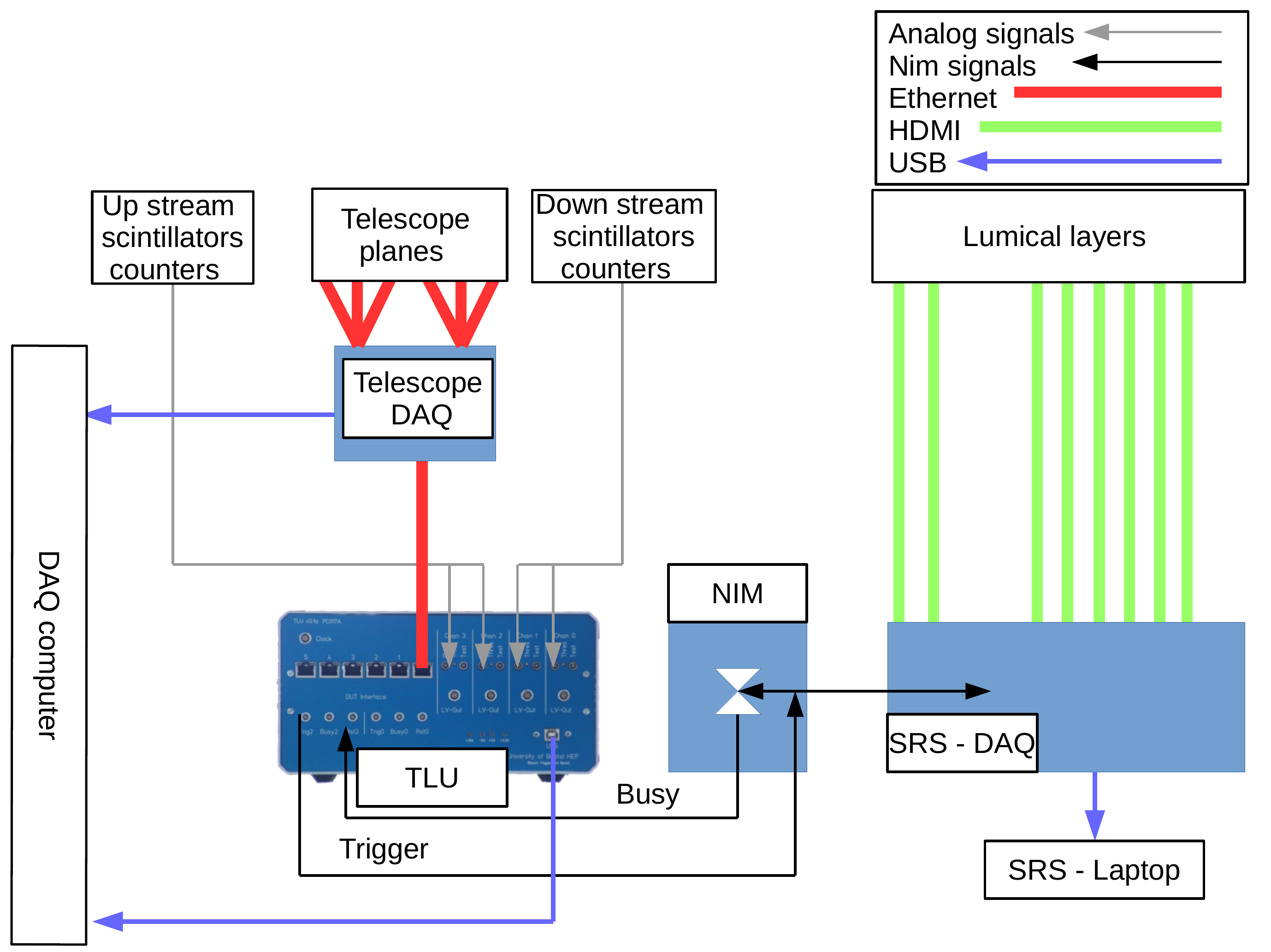}
  \caption{The Data Acquisition System.}
  \label{fig_tb2016_daq}
\end{figure}

A sketch of the Data Acquisition System (DAQ) is shown in Figure~\ref{fig_tb2016_daq}. 
It comprises two interdependent systems.
The first one is the EUDAQ which controls the beam Telescope and the Trigger Logic Unit, TLU. The second, the
calorimeter DAQ, is based on the Scalable Readout System (SRS)~\cite{SRS}, developed by the RD51 collaboration, and described below.  
A trigger signal is generated in the TLU, as a coincidence of signals from the scintillator counters Sc1 and Sc2, both consisting of two thin scintillators 
with attached photomultipliers.
The TLU then sends the trigger signal to both the Telescope acquisition and to the SRS. In addition, a BUSY signal is provided by a NIM logic
to prevent the TLU from sending more signals before the event acquisition ended.
The SRS, with a front-end hybrid board~\cite{APV_Hybrid} based on the APV25 front-end chip, is used for the readout. The APV25 front-end board has 128 readout channels, each consisting of a charge sensitive preamplifier and a shaper with a CR-RC filter producing a 50~ns shaped voltage pulse~\cite{APV_ieee,APV_nima}. The output of the shaper is sampled at 40 MHz and stored in an analog pipeline. During the beam test, the APV25 front-end boards are configured to operate in multi-mode, transmitting, upon receipt of a trigger from the TLU, 21 consecutive pipeline samples of each channel to the adapter board of the SRS through 3-m long HDMI cables. These samples are converted to 12-bit numbers in the SRS adapter board and transmitted to the data acquisition PC. 

Simulation results for the present configuration show that a single pad in a shower can be hit by 80 relativistic particles, hereafter referred to as MIPs ~\ref{fig_si_sensor}(see Figure~\ref{fig_pad_energy_depos} in Section~\ref{Calibration_APV_25}). The usage of the APV25 front-end board, which has a dynamic 
range for energy depositions originating from up to 8 MIPs, is hence not appropriate to read out sensor pads inside an electromagnetic shower. In order to enable measurements of a wider range of deposited energies, a capacitive charge divider is connected to the input of the APV25 front-end board. The attenuation factor of the charge divider is optimised by using the results from MC simulation. However, small signals from pads with low energy depositions in the tails of the shower are then below the detection threshold. The simulation of the observed noise level and the geometry of the present calorimeter   shows that an attenuation of the signal with a factor of 3.5~--~4.5, results in a~5~--~7\% loss of the deposited energy, which can be corrected for as described in Section~\ref{Calibration_APV_25}.

%

\section{Signal Processing}
\label{Signal_processing}
\begin{figure}[h!]
  \begin{minipage}[t]{0.48\textwidth}
    \includegraphics[width=\textwidth]{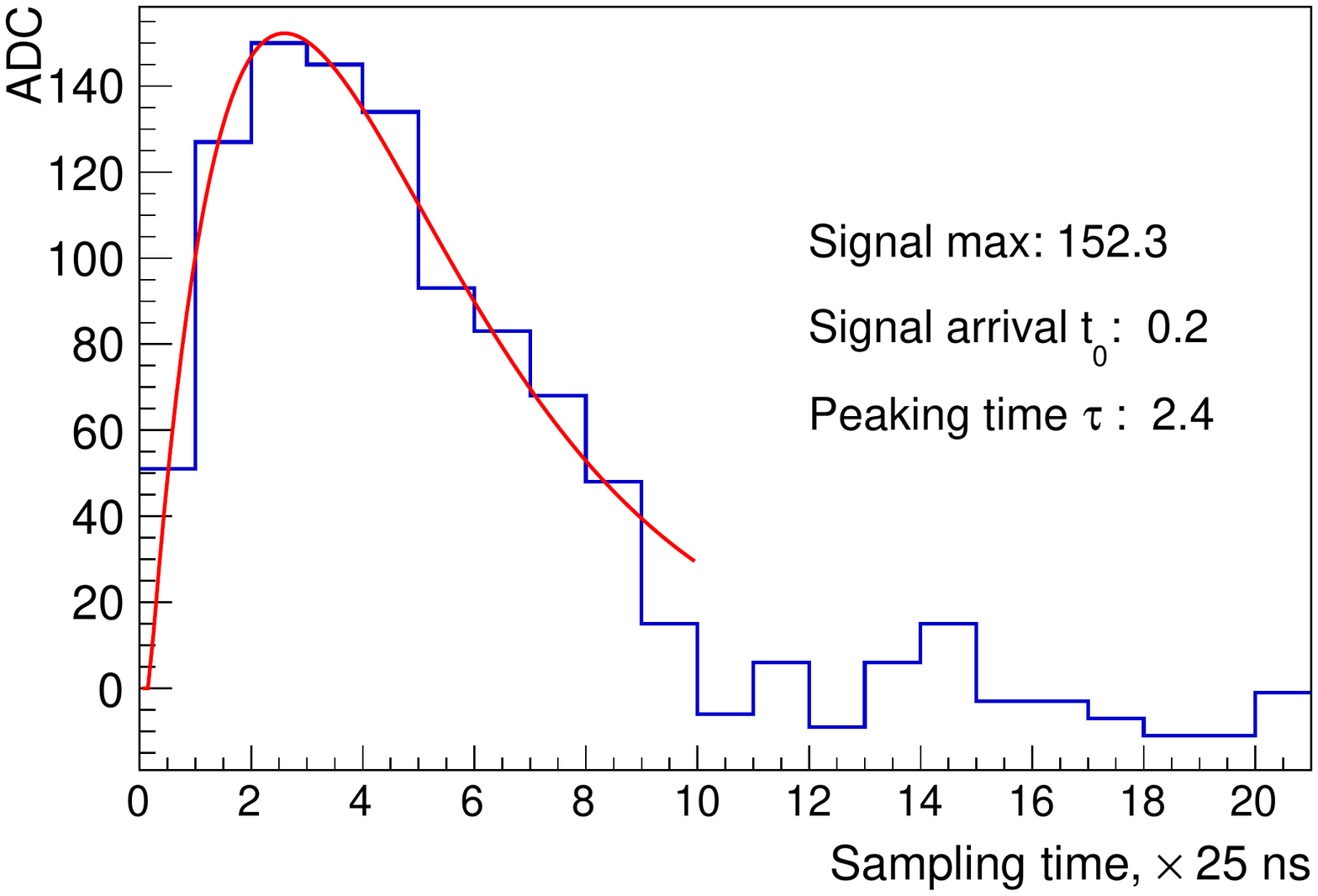}
    \caption{Signal waveform sampled in 25 ns time intervals. The red line is a fit with CR-RC-response function  of eqn.(\ref{eq_signal_t0_tau}).}
    \label{fig_raw_signal}
  \end{minipage}\hfill
  \hspace{0.03\textwidth}
  \begin{minipage}[t]{0.48\textwidth}
    \includegraphics[width=\textwidth]{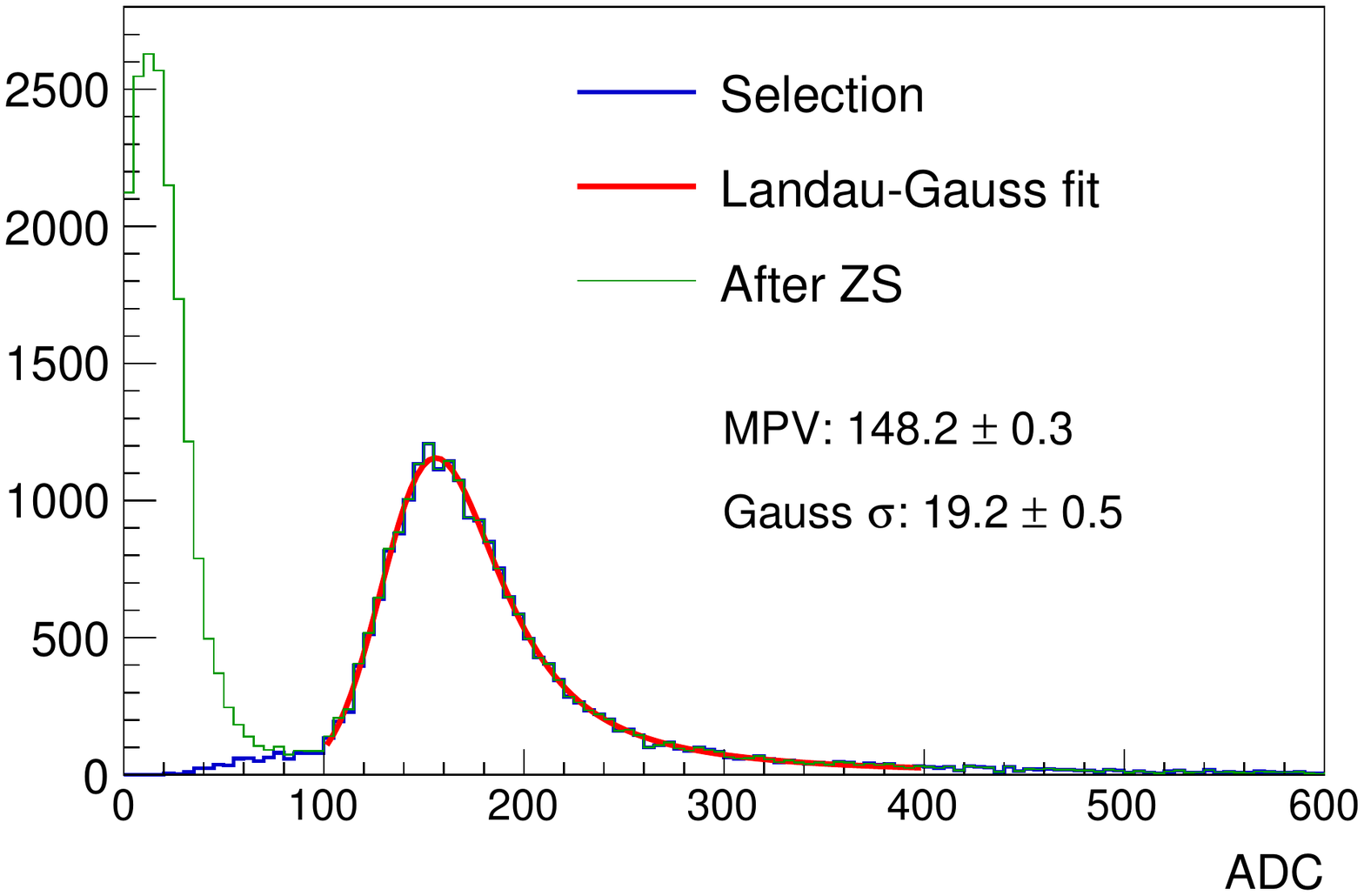}
    \caption{Signal distribution in a single pad of the tracking layer. Green line~-- after zero suppression, blue line~-- after additional selection criteria, and red line -- fit with a convolution of Landau and Gaussian distribution functions.}
    \label{fig_tracker_en_distrib}
  \end{minipage}
\end{figure}

The APV25 front-end chip operating in the multi-mode provides readout of 21 consecutive pipeline samples. The baseline of the output for each channel is calculated as the average 
of these samples in a dedicated pedestal run with a random trigger without beam. The noise is estimated as the standard deviation of the samples in the pedestal run and is used for setting the threshold in data during the run with a beam. An example of the signal for a single channel, after baseline and common-mode noise subtraction, is shown in Figure \ref{fig_raw_signal}. During data taking, the average of 21 samples of each channel is calculated and compared to the zero suppression (ZS) threshold. If it is below the ZS threshold, the data for the channel is not recorded. The threshold is set to~0.4 times the channel noise which results in a low enough threshold not to reject the signal from particles. 
Data is collected asynchronously, i.e. the readout electronics is not synchronised with the accelerator clock. As a consequence, most of the time the signal is not sampled exactly at its maximum. To determine the signal maximum, the samples are fitted with a CR-RC filter response function, as shown in Figure~\ref{fig_raw_signal}, 
\begin{equation}{
   S\left(t\right) = A\frac{t-t_{0}}{\tau} e^{-\frac{t-t_{0}}{\tau}} \Theta\left(t-t_{0}\right) ,
  }\label{eq_signal_t0_tau} 
\end{equation}
where $t_{0}$ is the arrival time of the signal, $\tau$~=~50~ns is the peaking time of the APV25 front-end board and $A$ is the relative signal amplitude. The function $\Theta\left(t-t_{0}\right)$ is the Heaviside step function. 

The relatively low ZS threshold allows a significant amount of noise pulses to pass through and further signal selection criteria are applied in the analysis. 
First, an artificial neural network (ANN) is used to analyse the signal and classify the data based on its shape. The ANN is represented by multilayer perceptron model with 21 inputs fed from the APV25 samples and one hidden layer with 10 nodes. The training set for different signal amplitudes is generated using the function in eqn.(\ref{eq_signal_t0_tau}) with a Gaussian noise added to each sample. 
After signal preselection based on the ANN, the signal is fitted with eqn.(\ref{eq_signal_t0_tau}), where the amplitude, arrival time~$t_{0}$ and peaking time~$\tau$ are used as parameters. To further improve the purity of the signal, selection criteria are applied to the parameters~$t_{0}$ and~$\tau$. The efficiency of the selection is studied using external pulses, as described in Section~\ref{Calibration_APV_25}. 
%

\begin{figure}[h!]
  \begin{minipage}[t]{0.48\textwidth}
    \includegraphics[width=\textwidth]{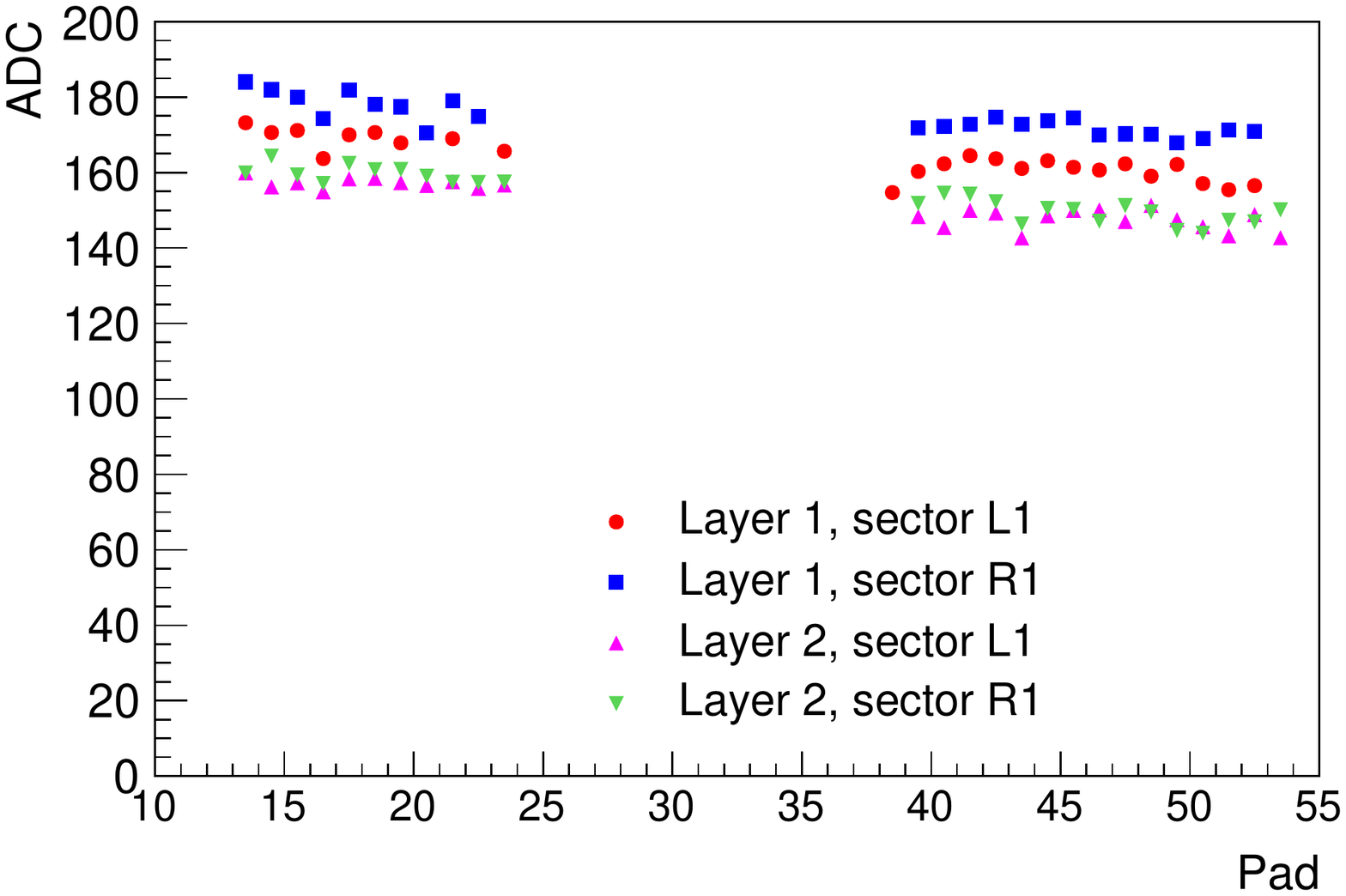}
    \caption{Most probable value of the signal in the pads of the tracking layers covered by the electron beam of 5~GeV.}
    \label{fig_mpv_all}
  \end{minipage}\hfill
  \hspace{0.03\textwidth}
  \begin{minipage}[t]{0.48\textwidth}
    \includegraphics[width=\textwidth]{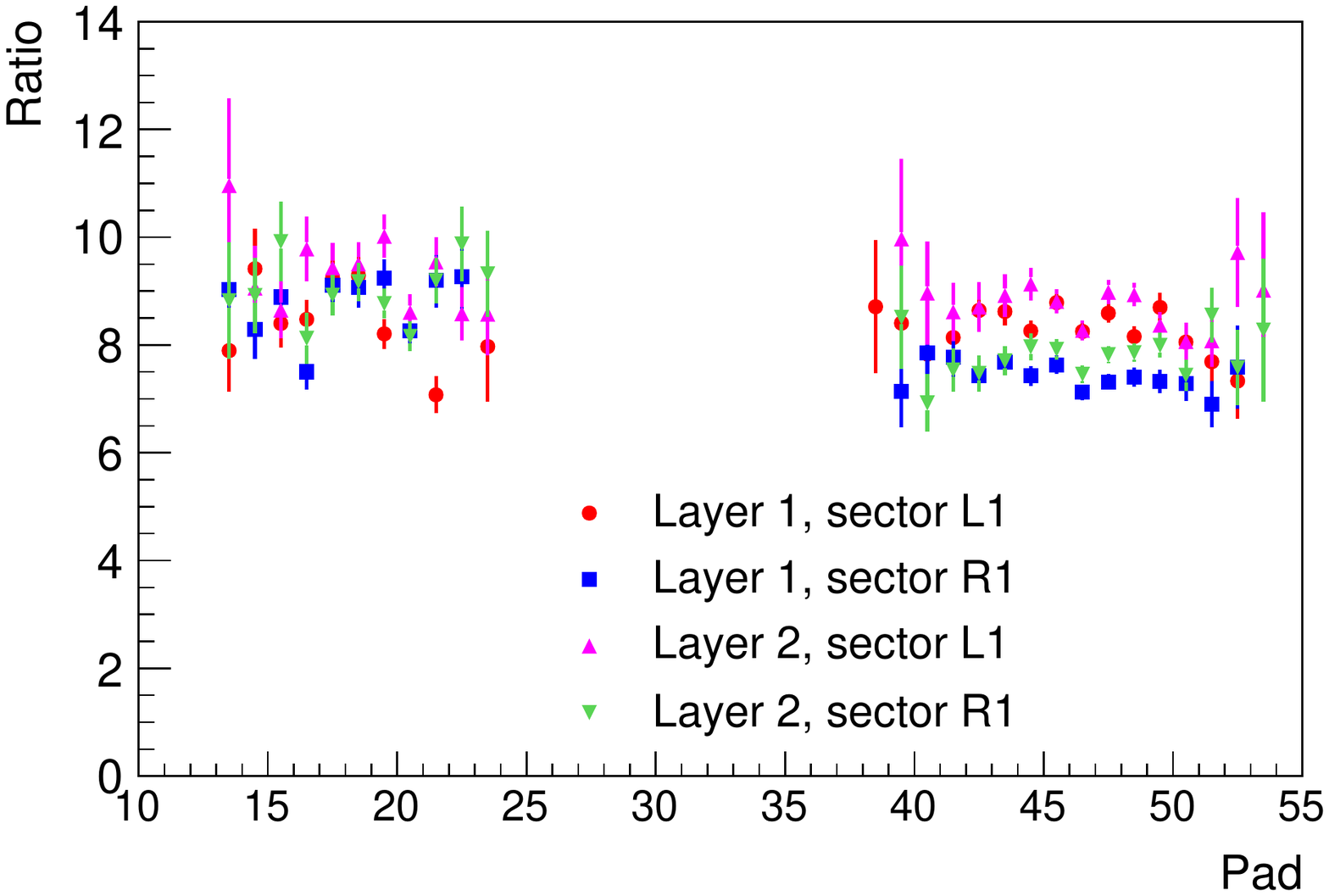}
    \caption{Signal to noise ratio for the pads of the tracking layers covered by the electron beam of 5~GeV.}
    \label{fig_sn_all}
  \end{minipage}
\end{figure}

Figure~\ref{fig_tracker_en_distrib} shows the distribution of the signal amplitudes produced by a 5~GeV electron beam and measured in a single channel of the tracking plane. The green line corresponds to the data which pass the ZS threshold. The blue line, which corresponds to the data after applying additional signal selection criteria, illustrates the effective noise suppression in the analysis. The most probable value (MPV) of the peak is estimated using a fit with a convolution of Landau and Gaussian distribution functions. The width
$\sigma$ of the Gaussian distribution is considered as noise measurement. The MPV values of the amplitude distribution corresponding to 5 GeV electrons are shown in Figure~\ref{fig_mpv_all}. The higher values for small pad numbers reflect the geometry of the sensor where these pads have smaller area and smaller capacitance. The same effect is observed for the signal-to-noise ratio shown in Figure~\ref{fig_sn_all}. Since the beam profile has blurry edges, the statistical uncertainties increase for pads that correspond to the periphery of the beam. For most of the channels the signal-to-noise ratio is within a range from 7 to 10.
The most probable value 
of the energy deposited by 5~GeV electrons is used to define the unit MIP for the energy deposition in the sensors. 
Based on MC simulations, a MIP corresponds to 88.5~keV.

\begin{figure}[h!]
\begin{center}
    \includegraphics[width=0.48\textwidth]{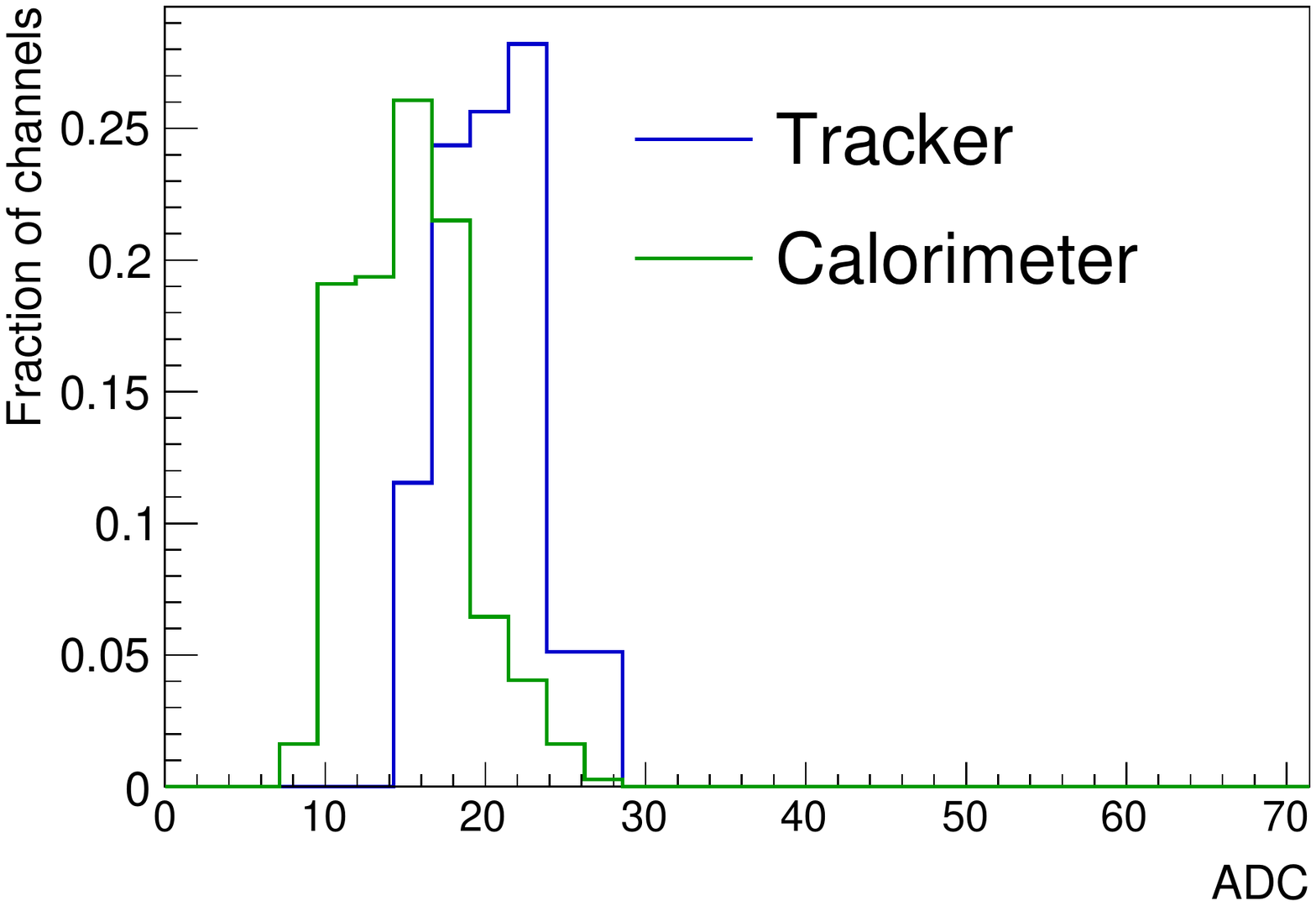}
    \caption{Noise distribution in the channels of the tracking layers (blue) and the calorimeter layers (green).}
    \label{fig_noise_track_calo}
  \end{center}
\end{figure}

For the detector planes that are installed in the calorimeter,   the capacitive charge divider is used. The signals from single particles
are too small to be registered, and hence the signal-to-noise ratio cannot be measured. Taking into account the design of the charge divider and the noise measured in the pedestal run, as shown in Figure~\ref{fig_noise_track_calo}, the estimated value of signal-to-noise ratio is in the range of 2~--~3. For such a low ratio, the signal-shape analysis, using ANN and selection-criteria for the parameters retrieved from the fit, allows for the efficient identification of the signal with little contamination from noise.

\section{Calibration of the APV25 front-end board}
\label{Calibration_APV_25}
The linearity of the APV25 front-end boards was studied with the bare chip~\cite{APV_ieee,APV_nima} and it was found to be very good for signals of up to 3~MIPs and remains better than~5\% up to 5 MIPs. 

The relative response of the APV25 channels, equipped with a capacitive charge divider, is measured
using a voltage pulse supplied to the channel input through a capacitor of 2~pF. The detector capacitance is simulated by a 7~pF capacitor connected in parallel to the channel input. About 10 randomly chosen channels for each APV25 front-end chip 
were measured and the average response curve was calculated for each APV25.

%
\begin{figure}[h!]
\begin{minipage}[t]{0.48\textwidth}
   \includegraphics[width=\textwidth]{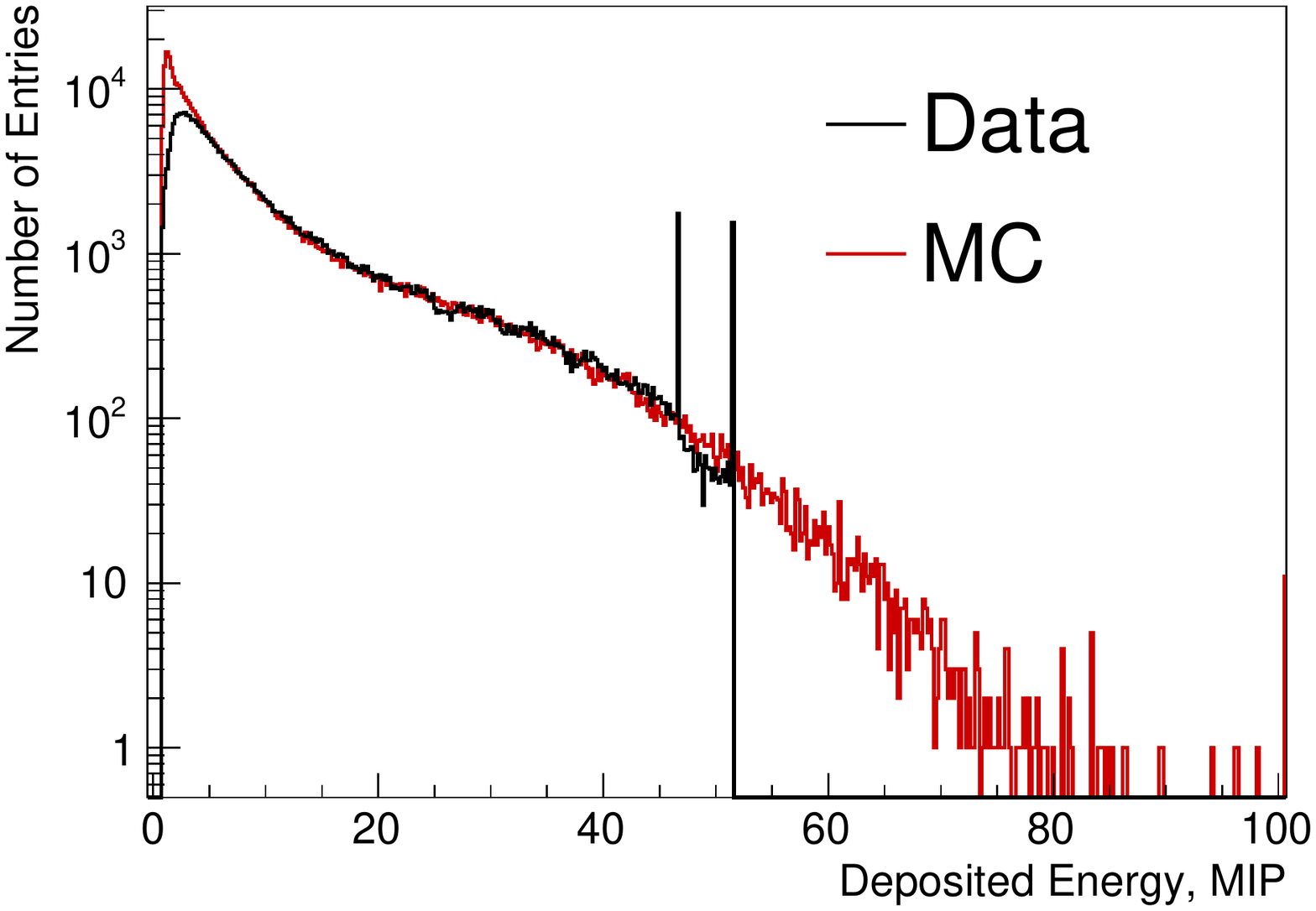}
    \caption{Distribution of the deposited energy in a sensor pad in the detector layer after 5 tungsten plates. The red line is a MC simulation, 
    and the black line is data using as calibration an interpolation between measured 
    calibration values of the APV25 front-end chip. The sharp spikes are due to saturation in two APV25 front-end boards.}
    \label{fig_pad_energy_depos}
\end{minipage}\hfill
\hspace{0.03\textwidth}
\begin{minipage}[t]{0.48\textwidth}
    \includegraphics[width=\textwidth]{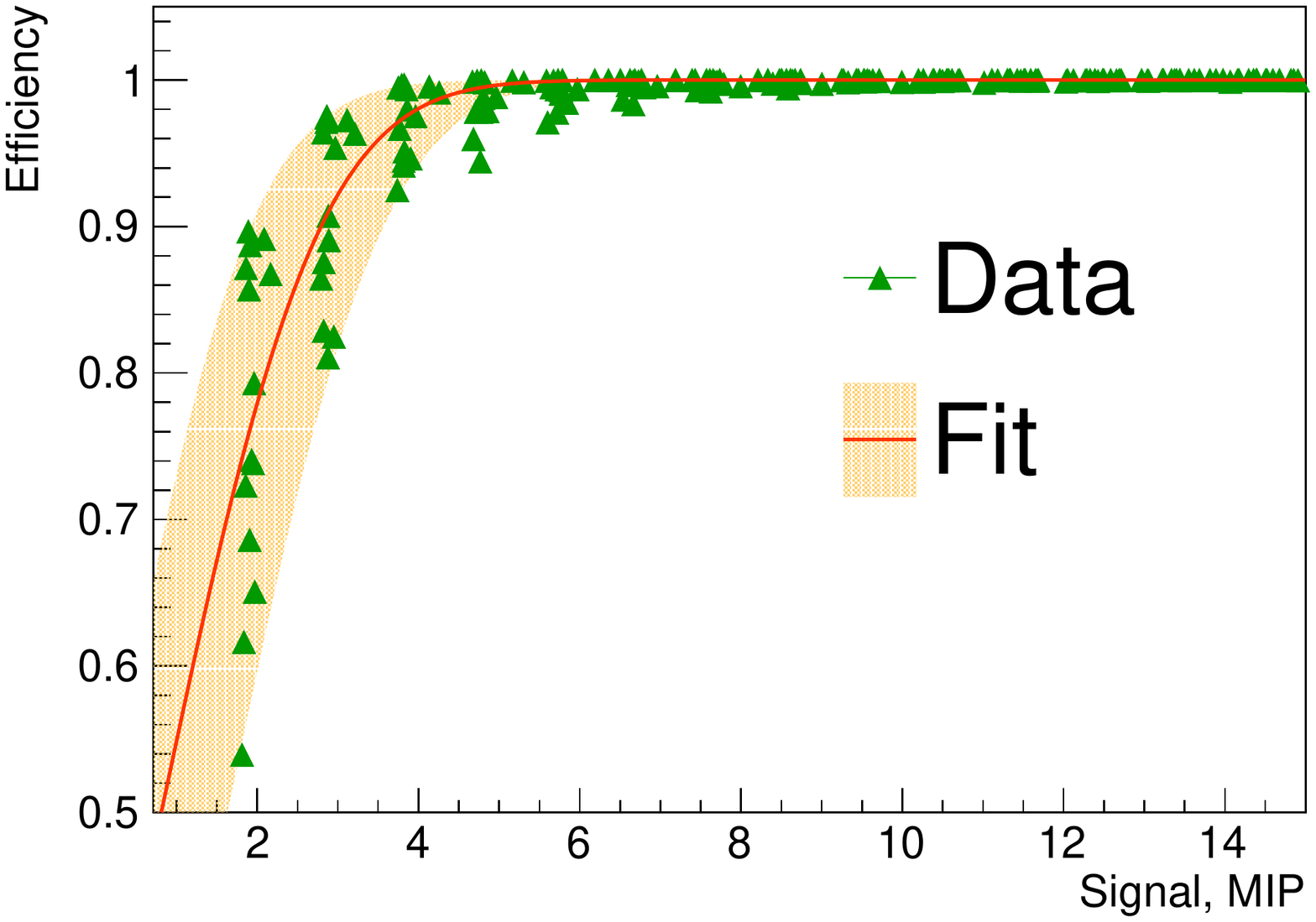}
    \caption{Efficiency of signal identification as a function of the signal amplitude. 
    Green triangles are measured for different channels, the red line is an average of the fit using eqn.~(\ref{eq_efficiency}) to a large number of channels, and the shaded area corresponds to the spread of fits at small amplitudes.}
    \label{fig_apv_efficiency}
\end{minipage}
\end{figure}

The APV25 front-end board with charge divider approaches saturation at about 1600~ADC counts. In this analysis, 
the maximum signal size is 1450~ADC counts, reasonably below the saturation.

Figure~\ref{fig_pad_energy_depos} shows the distribution of the deposited energy in a pad in the detector layer after 5 tungsten plates. The data were processed with the calibration 
obtained by interpolation between measured values. 
The sharp spikes are due to saturation which, after calibration, has slightly different thresholds for each APV25. 
The measured distribution of the deposited energy in a single pad is well reproduced by the simulations for signal amplitudes larger than 5 MIPs.
However, smaller signals become masked by the noise. This loss of signals can also be seen in Figure~\ref{fig_pad_energy_depos}, where for small amplitudes
the experimental distribution is below the MC 
expectation. 

In order to correct for this loss of signals, the efficiency $\epsilon$ of identifying the signal of a small amplitude is studied with the same setup using an external voltage pulse.
We define the efficiency of signal identification as the ratio of the number of identified signals to the number of generated ones.
This ratio depends on the signal-to-noise ratio and therefore is slightly different for different APV25 front-end chips, as shown in Figure~\ref{fig_apv_efficiency} where the results for channels of different APV25 front-end chips are presented. 
For each APV25 front-end chip, about 10 channels are measured. For signals larger than 10 MIPs, the efficiency is 100\% in all channels. 
For a smaller number of MIPs, some channels give lower efficiencies. The measurements of the efficiency $\epsilon$ are fit by the following expression:
\begin{equation}{
   \epsilon = p_{0}\left(1+\mbox{\textit{erf}}\left( \frac{S-S_{0}}{p_{1}} \right) \right)
  }\label{eq_efficiency} 
\end{equation}
where \mbox{\textit{erf}} is the error function, $S$ the signal amplitude and $p_{0}$, $p_{1}$ and $S_{0}$ are fit parameters. 
The red curve in Figure~\ref{fig_apv_efficiency} represents the average of the fit of a large number of channels and 
the shaded area the spread of the fit in these channels at low signal amplitudes. 
Since the noise level observed during lab calibration measurements and beam test are similar, the efficiency correction for small 
signal sizes is applied to the test-beam simulations using the results of the fit to eqn.~(\ref{eq_efficiency}).

\section{Results}
\label{results}

More than seven million events were collected 
in an electron beam from 1~GeV to 5~GeV energy, with 1~GeV steps, for different setup configurations to 
measure the precision of the shower position determination, the electromagnetic shower development in longitudinal and transverse directions and the effective Moli\`ere radius. 

%
\begin{figure}[h!]
  \begin{minipage}[t]{0.45\textwidth}
    \includegraphics[width=\textwidth]{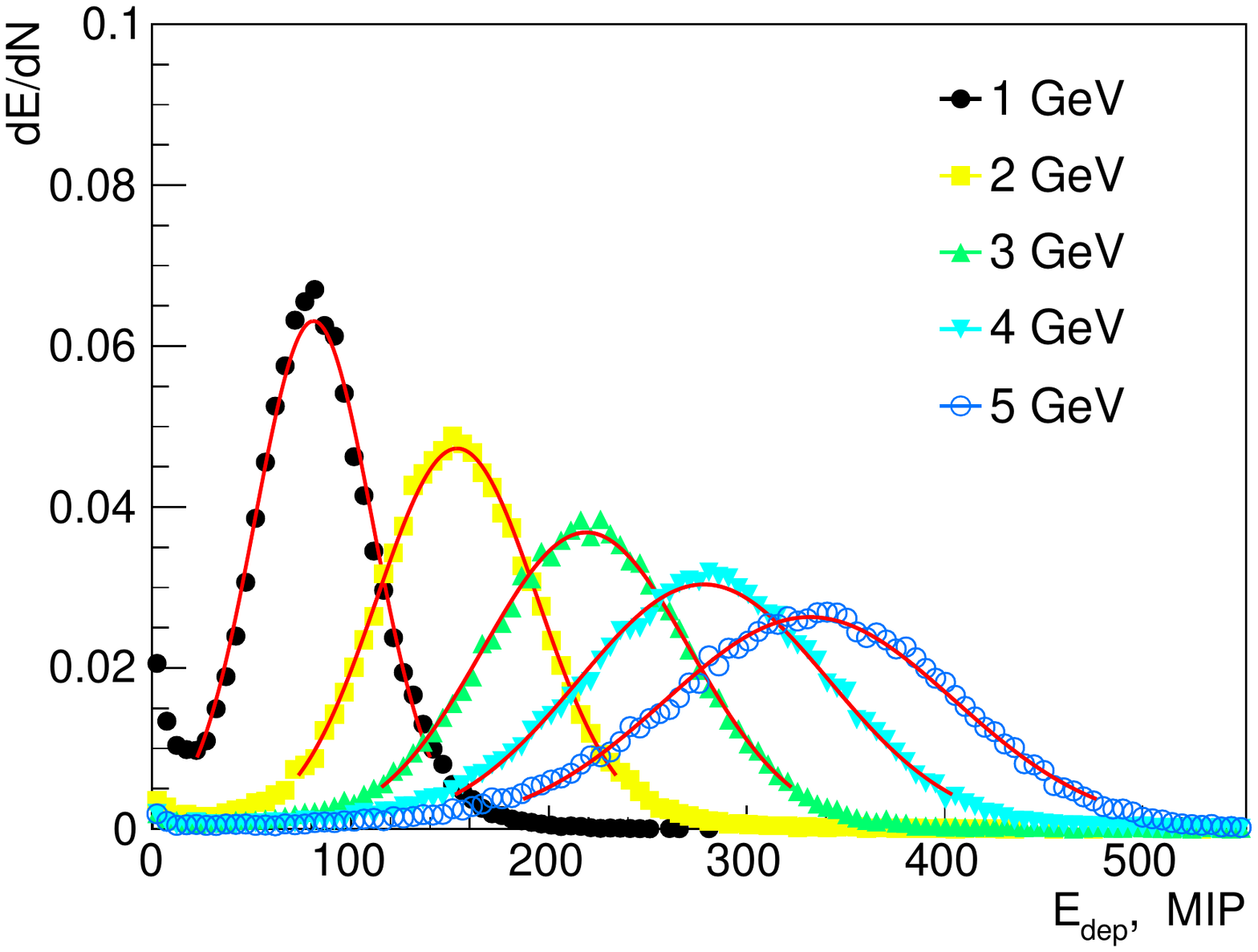}
    \caption{Distribution of the energy deposited in the calorimeter, $dE/dN$, for different electron energies. The lines illustrate fits with a Gaussian.}
    \label{E_dep_total_1_5gev}
  \end{minipage}\hfill
  \hspace{0.025\textwidth}
  \begin{minipage}[t]{0.45\textwidth}
    \includegraphics[width=\textwidth]{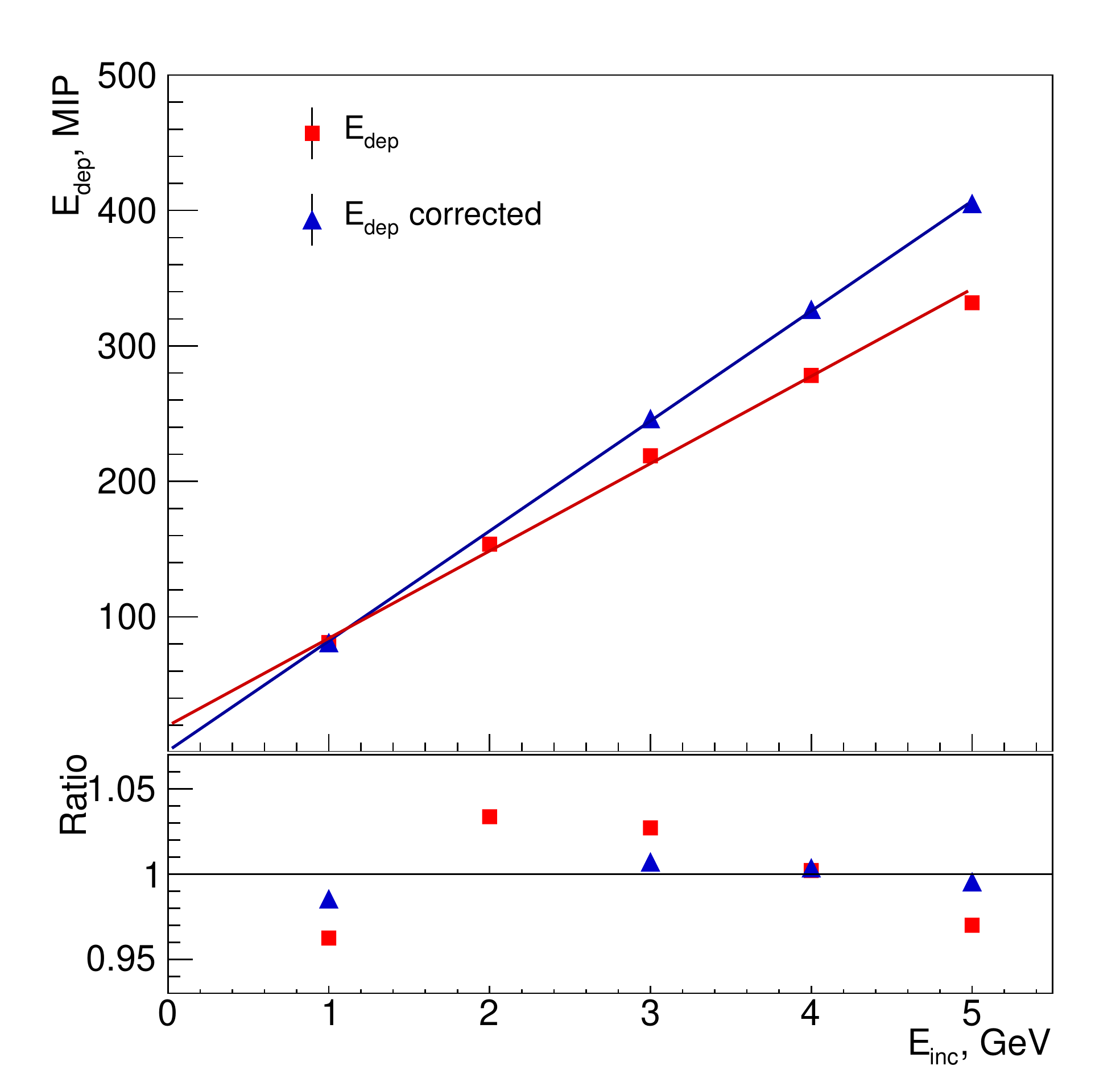}
    \caption{Average deposited energy in the calorimeter, $E_{dep}$, as a function of beam energy before (red) and after applying the APV25 calibration and corrections to the leakage fraction estimated from the simulation (blue). The lines are straight line fits to the data. The lower part of the figure shows the ratio of the deposited energy to the straight line fit.}
    \label{Avarage_E_dep_total_1_5gev}
  \end{minipage}
\end{figure}

Figure~\ref{E_dep_total_1_5gev} shows the distributions of the energy deposited in the sensors of the calorimeter 
for beam electrons of different energy. The average deposited energy as a function of the electron beam energy is presented in Figure~\ref{Avarage_E_dep_total_1_5gev}. The measured raw values increase with increasing beam energy, with a tendency of a reduced slope at larger beam energies. After applying the APV25 calibration, as described in Section~\ref{Calibration_APV_25}, and correcting for the energy leakage fraction, estimated from the simulation, the response becomes nearly linear.

\subsection{Reconstruction of the shower position}


For the reconstruction of the shower position, pads with deposited energy
were combined into clusters. In the first step, the depositions in all pads at a given radial and azimuthal position are summed over all detector layers. 
The clustering algorithm used in this study builds a cluster including all nearest neighbour pads. 
The pad with radial number~$n$ and sector number~$k$ is assigned to a cluster if the cluster contains a pad with radial number~$n'$ and sector numbers~$k'$ such that both $|n-n'| \leq 1$ and $|k-k'| \leq 1$. If this holds, the cluster is considered as an electromagnetic shower.
The shower position is determined using a weighted sum:
\begin{equation} \label{eq:cog}
Y_c = \frac{\sum\limits_{m} Y_{m} w_m}{\sum\limits_{m} w_m},
\end{equation}
where the index $m$ runs over all pads included in the shower. $Y_{m}$ is the position of the pad and $w_m$~is a weight, which in the simplest approach could be taken as the energy $E_m$ deposited in the pad. It has however been shown~\cite{double_exp_cher,double_exp_reco,awes_position_ln}  that this approach gives a biased estimate when the shower position is not in the centre of a pad. Several methods were developed to achieve more accurate position reconstruction, and the following choice of  weights is found to be the most appropriate:
\begin{equation} \label{eq:log_weight}
w_m = max\left\{0; W_0+\ln{\frac{E_m}{\sum\limits_{j} E_j}}\right\},
\end{equation}
where $W_0$ is a free dimensionless parameter. The performance of the clustering algorithm is studied in a simulation and for the present configuration the best resolution for the radial coordinate of the shower is achieved with $W_0=3.4$. 

\begin{figure}[h!]
  \begin{center}
    \includegraphics[width=0.6\textwidth]{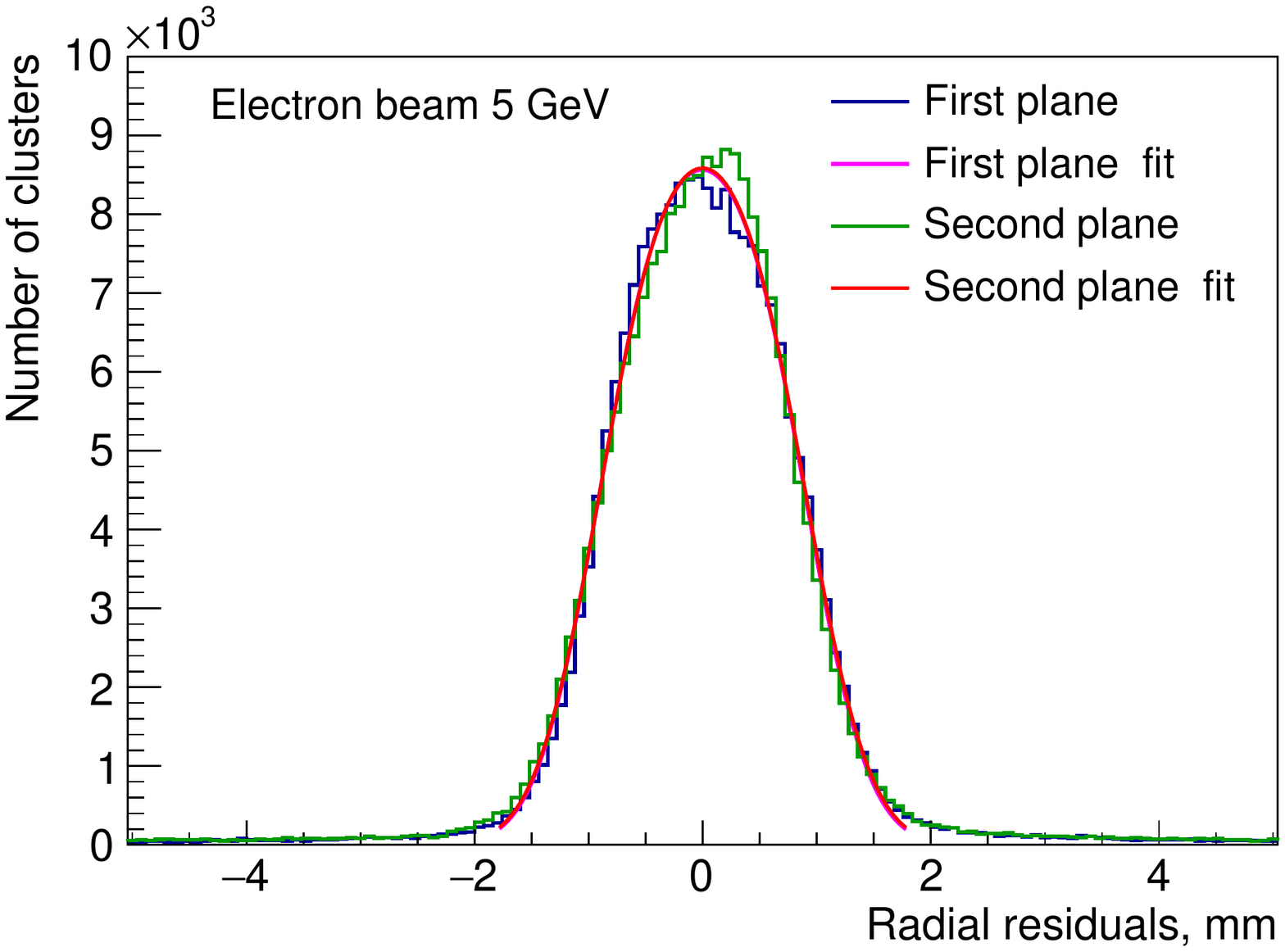}
  \end{center}
  \caption{Distribution of residuals of the radial position measurements in the tracking planes and the calorimeter, obtained with a 5 GeV electron beam. 
  }
  \label{fig_tracker_lumical_residuals}
\end{figure}

The resolution of the shower position reconstruction in the calorimeter is estimated using the tracker planes. Two detector planes are installed at distances of 86~mm and 63~mm in front of the first tungsten plate. Because of the relatively large pad size, about~95\% of the reconstructed clusters in the tracking planes consist of one pad, hence charge sharing between pads cannot be used for the position reconstruction. The impact position of beam particles is set to the middle of the pad. Since the beam particle density is found to be almost constant, a uniform distribution of beam particles within the pitch of the sensor is given. Assuming that the uncertainty of the shower position reconstruction in the calorimeter has a Gaussian distribution, the distributions of the residuals between the particle position in the tracking plane and in the calorimeter is described by the convolution  
\begin{equation} \label{eq_residuals_fit_function}
f(x) = \frac{B}{p\sigma \sqrt{2\pi }} \int\limits_{x_{0}-\frac{p}{2}}^{x_{0}+\frac{p}{2}}  e^{-\frac{(x-z)^2}{2\sigma ^2}} \mathrm{d}z,
\end{equation}
where $\sigma$ is the position resolution in the calorimeter, $p$ the pitch of the tracking plane, $x_{0}$ accounts for relative displacement and $B$ provides the normalisation for a given number of events.
Figure~\ref{fig_tracker_lumical_residuals} shows the distribution of the residuals of the reconstructed radial position of the shower in the calorimeter and in the two planes of the tracker. To test the performance of the method the pitch of the sensor can be also considered as a fit parameter. In this case the values found from the fit are~1.86~mm and~1.71~mm for the first and second tracking planes, respectively. These numbers are within 5\% equal to the sensor pitch of~1.8~mm. The resolution $\sigma$ of the shower position reconstruction, found from the fit when $p$~is fixed to the value of the sensor pitch, is (440 $\pm$ 20)~$\mu$m, and the absolute values of relative displacements $x_{0}$ are less than~5~$\mu$m. 

The small distortion seen at the top part of the distribution for the second tracker plane in Figure~\ref{fig_tracker_lumical_residuals} is explained by the small asymmetry of the beam profile and circular geometry of the sensor which, in combination, result in a decline from the uniform distribution of the position uncertainty in the tracking planes.

\subsection{One dimensional transverse shower profile}

The one dimensional profile of the deposited energy in the sensor layers for each event is obtained as the following sum: 
\begin{equation} \label{eq_E_pad_layer}
E^{det}_{nl} = \sum\limits_{k} \epsilon_{nkl} \ ,
\end{equation}
where $\epsilon_{nkl}$ is the deposited energy measured in the sensor pad with radial number~$n$, sector~$k$ and layer~$l$. The sector index~$k$ runs over two central sectors of the sensor considered and the layer index~$l$ corresponds to the 5 detector planes of the calorimeter. 
About 5\% of randomly distributed channels in the calorimeter have a larger noise level corresponding to signal sizes of up to 40 MIPs. 
The influence of these channels, hereafter referred to as bad channels, on the shower development study is eliminated by calculating $\langle E^{det}_{nl} \rangle$ for all indexes $n$ and $l$ only from properly working channels.

\begin{figure}[h!]
  \begin{minipage}[t]{0.31\textwidth}
    \includegraphics[width=\textwidth]{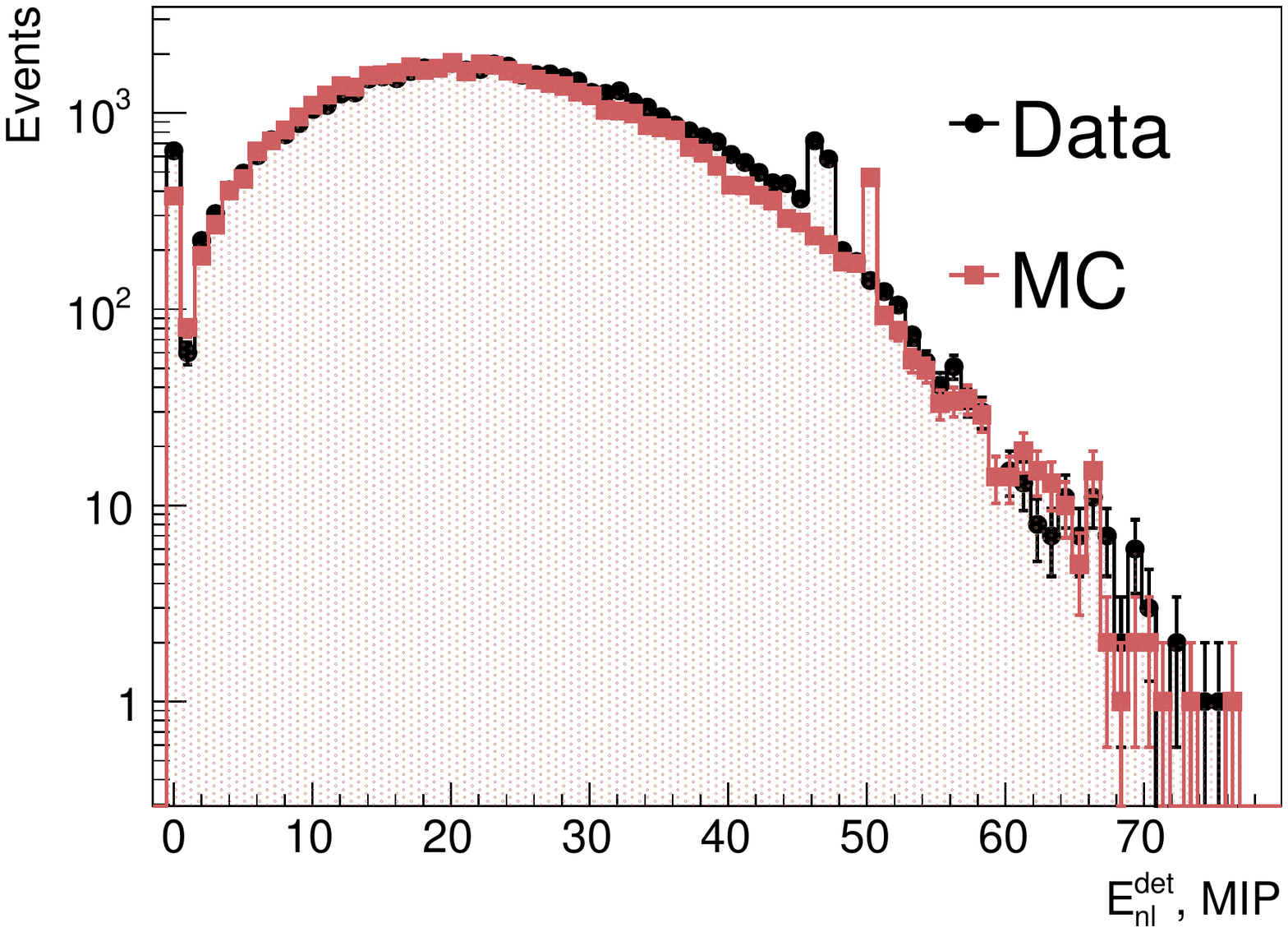}
    \center a)
  \end{minipage}\hfill
  \hspace{0.02\textwidth}
  \begin{minipage}[t]{0.31\textwidth}
    \includegraphics[width=\textwidth]{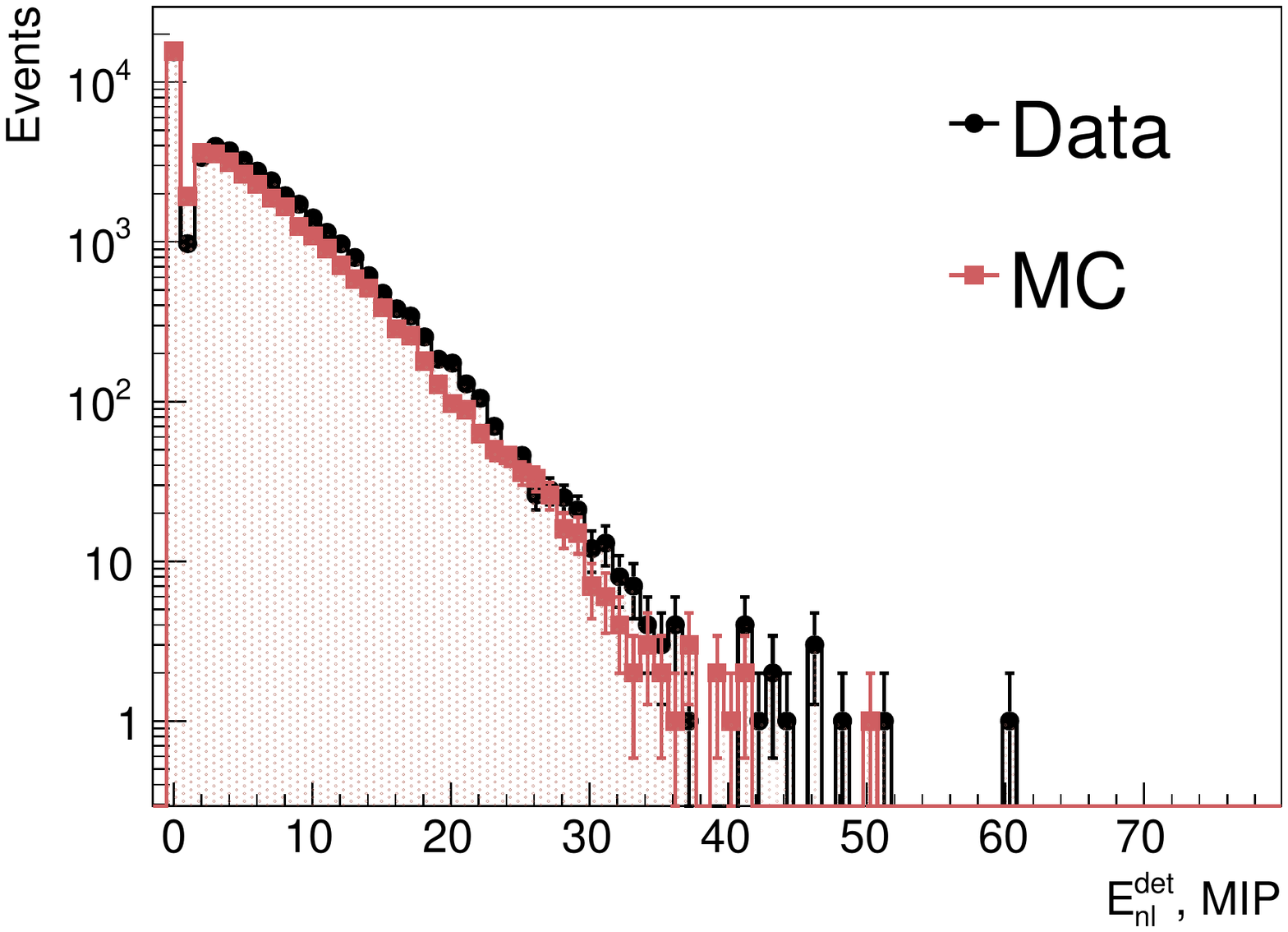}
    \center b)
  \end{minipage}
  \hspace{0.02\textwidth}
  \begin{minipage}[t]{0.31\textwidth}
    \includegraphics[width=\textwidth]{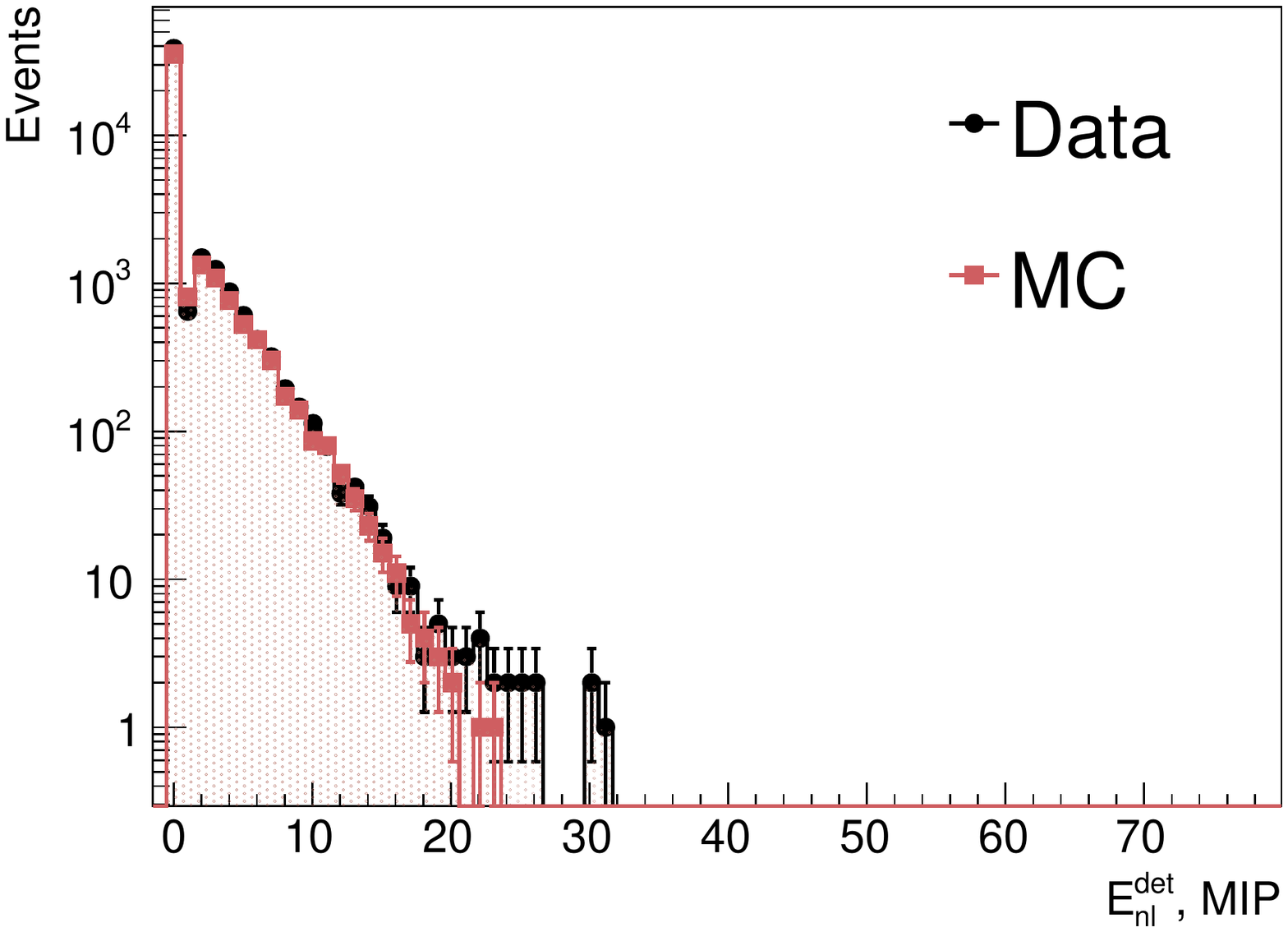}
    \center c)
  \end{minipage}
  \caption{Distributions of deposited energy $E^{det}_{nl}$ as defined in eqn.~(\ref{eq_E_pad_layer}) summed over two sensor sectors L1 and R1 in a layer after seven tungsten plates for radial pads which correspond to shower core~(a), two pads away from the core~(b) and five pads away from the core~(c). The distributions are obtained with a 5 GeV electron beam.}
  \label{fig_e_dep_layer_pads}
\end{figure}

Since the particle position changes from event to event due to the transverse beam size within about 10 pads, for the estimation of the average value of $\langle E^{det}_{nl} \rangle$, the index~$n$ in each event is set to n=0 for the pad which contains the centre of the shower. An example of the distributions of~$E^{det}_{nl}$ for the shower core ($n=0$) and pads with~$n=-2$, and~$n=-5$ for the layer after seven tungsten plates are shown in Figure~\ref{fig_e_dep_layer_pads}. Data is well described by the simulation.

\subsection{Longitudinal shower profile}

 The average energy~$\langle E^{layer}_{l} \rangle$ deposited in calorimeter layer~$l$ is calculated as the following sum: 
\begin{equation} \label{eq_E_longitudinal}
\langle E^{layer}_{l} \rangle = \sum\limits_{n} \langle E^{det}_{nl} \rangle \ ,
\end{equation}
where~$n$ runs over the radial pads of the two central sectors of the sensor. 
About 5\% of randomly distributed channels in the calorimeter have a larger noise level corresponding to signal sizes of up to 40 MIPs. 

\begin{figure}[h!]
  \begin{minipage}[t]{0.45\textwidth}
    \includegraphics[width=\textwidth]{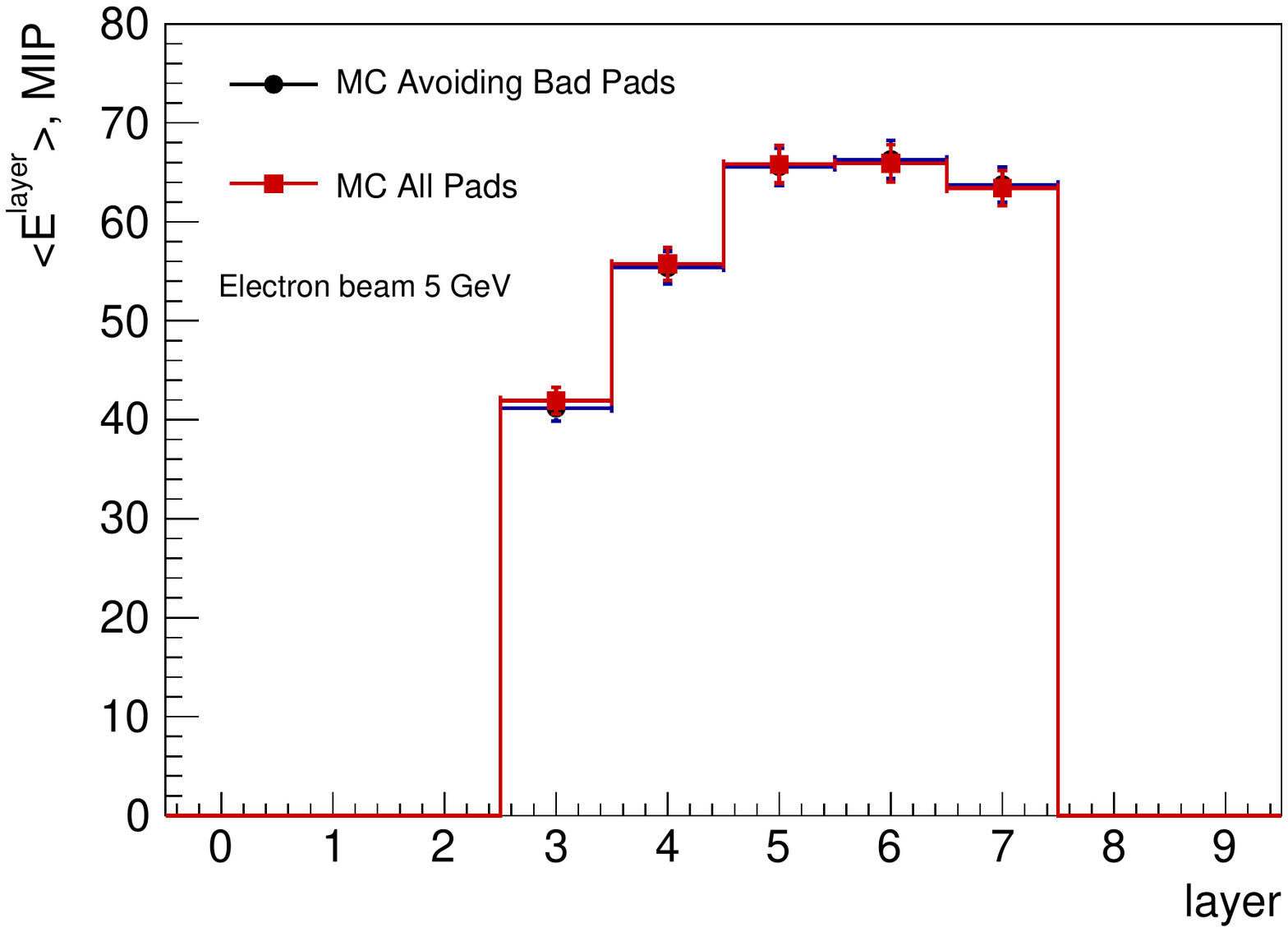}
    \caption{Longitudinal shower, comparison between simulations with and without bad channels. The distributions are obtained with a 5 GeV electron beam.}
    \label{fig_Longitudinal_mca}
  \end{minipage}\hfill
  \hspace{0.025\textwidth}
  \begin{minipage}[t]{0.45\textwidth}
    \includegraphics[width=\textwidth]{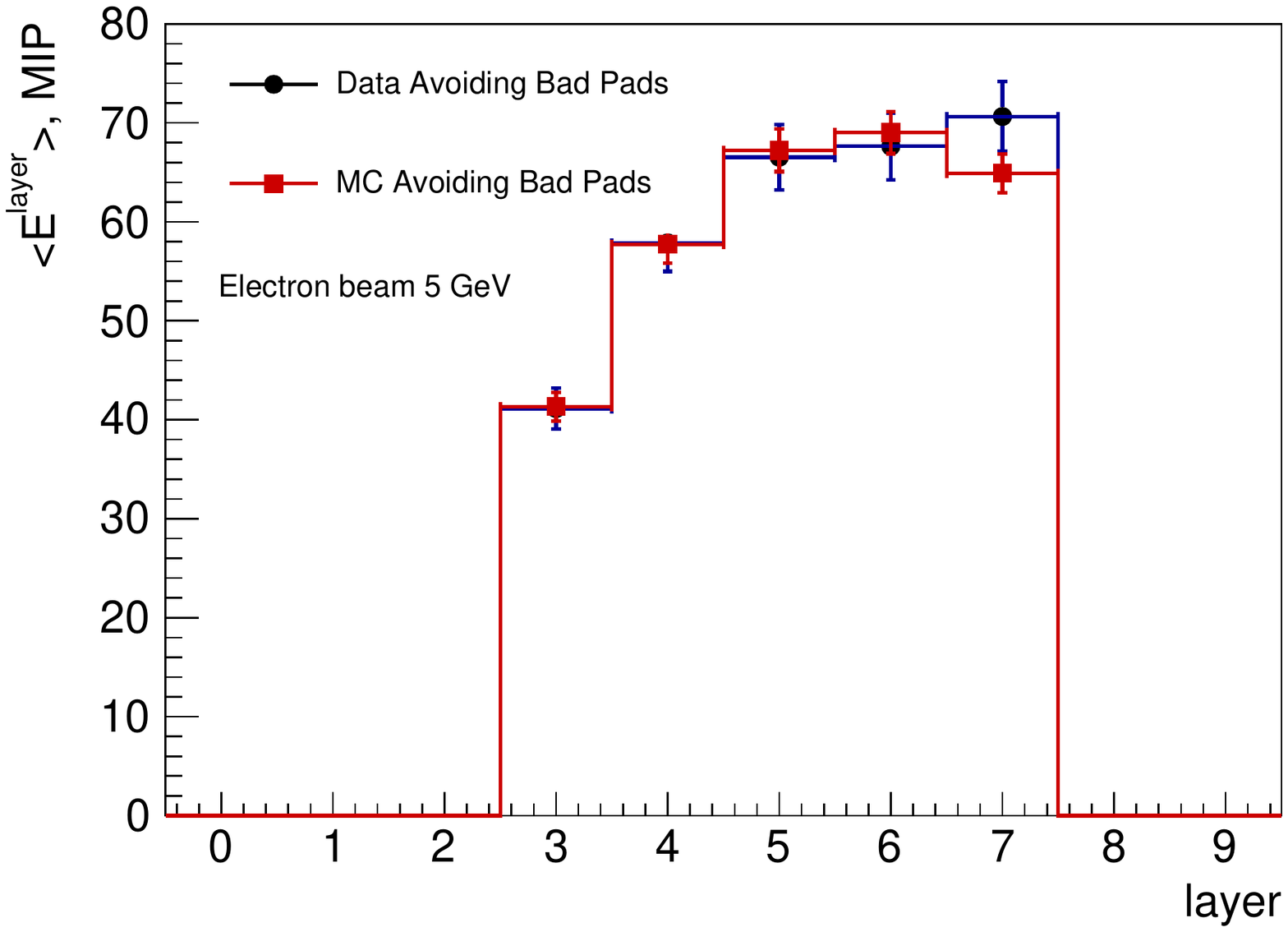}
    \caption{Longitudinal shower profile, comparison between data and simulation. The distributions are obtained with a 5 GeV electron beam.}
    \label{fig_Longitudinal_dmc}
  \end{minipage}
\end{figure}

A Monte Carlo simulation has been done to estimate the impact of bad channels on the longitudinal shower profile. The result is shown
in
Figure~\ref{fig_Longitudinal_mca}.
The red distribution corresponds to a calorimeter without bad channels and the black one is obtained after dropping bad channels, introduced in the simulation in the same locations as observed in data. Both distributions agree very well within statistical uncertainties.

The development of the longitudinal shower profile is then measured using only events with properly working channels. 
In Figure~\ref{fig_Longitudinal_dmc} the deposited energy as a function of the layer $l$ is shown for data and Monte Carlo simulation.
The maximum of the shower is reached in data at layer 7. Both distributions are, within statistical uncertainties, in reasonable agreement.


\subsection{The Moli\`ere Radius}
\label{RM_Measurement}

The sensor in Figure~\ref{fig_si_sensor} has a relatively fine segmentation in the radial direction, with a pitch of 1.8~mm, but the size of the sectors is between 2 and 2.5~cm in the irradiated area. Such a geometry does not allow to uniformly sample the electromagnetic shower in the transverse plane and requires the development of a dedicated method to measure the effective Moli\`ere radius. Such a  method was developed and presented in detail in Ref.~\cite{LumiCal_multilayer_tb2014_epjc}. Since here the same sensors are used, this method will be applied. 
It is briefly described in the following with small modifications which are mainly motivated by the difference in the design of the detector  plane.

Denoting $F_{E}( {r})$ the density function of the average deposited energy in the transverse plane with respect to the shower axis, the energy in the area covered by a single detector pad can be expressed as the integral
\begin{equation} \label{eq_pad_energy_int_s}
E_{n} = \int\limits_{S_{n}} \! F_{E} ( {r}) \, \mathrm{d}S,
\end{equation}
where $S_{n}$~is the area which corresponds to the sensor pad $n$. The function $F_{E}( {r})$ is cylindrically symmetric with respect to the shower axis, and is expressed in cylindrical coordinates with the origin at the center of the shower. Hence it depends only on the radius $r$. Since, on average, 90\% of the deposited energy lies inside a cylinder with a radius of one Moli\`ere radius $R_{\mathcal{M}}$, the following equation can be used for the Moli\`ere radius calculation:
\begin{equation} \label{eq_MR_1}
0.9 = \frac{\int\limits_{0}^{2\pi } d\varphi \int\limits_{0}^{R_\mathcal{M}}  \! F_{E} ( r ) r \, \mathrm{d}r}
     {\int\limits_{0}^{2\pi } d\varphi \int\limits_{0}^{\infty}  \! F_{E} ( r ) r \, \mathrm{d}r }  .
\end{equation}

The values of $E_{n}$ can be calculated using a parameterised trial functions~$F_{E}( {r})$. Fitting this trial function to the average deposited energy measured in the corresponding pads, one can define their parameters and use them in eqn.~(\ref{eq_MR_1}) to obtain the Moli\`ere radius. 

In the previous paper~\cite{LumiCal_multilayer_tb2014_epjc}, the circular shape of the pads was approximated for simplicity by a straight strip. The effect of this approximation was studied in a simulation~\cite{LumiCal_multilayer_tb2014_proc} and it was shown that the difference between values of $E_{n}$ calculated for pads of circular shape and for strip-like pads depends on the pad position with respect to the shower centre amounts to at most 2\%. This difference was included in the systematic uncertainty. This effect was also diminished in the data analysis because the detector  planes had limited number of pads connected to the readout and some values of $E_{n}$ could not be measured directly, but were recovered assuming the symmetry with respect to the shower core.

In the present study, the numerical integration in eqn. (\ref{eq_pad_energy_int_s}) is done using the correct geometry of the sensor pad. 
To this end it is convenient to use cylinder coordinates which are linked to the sensor geometry. Changing the coordinates to $\vec{r} =  \vec{r'} -  \vec{R'_{0}}$, where 
$\vec{R'_{0}}$~is the position of the shower axis in the sensor reference frame, the pad energy can be obtained by the integration:
\begin{equation} \label{eq_pad_energy_int_rphi}
E_{n} = \int\limits_{\varphi'_{min}}^{\varphi'_{max}} \int\limits_{r'_{n}}^{r'_{n+1}} \! F_{E} (| \vec{r'} -  \vec{R'_{0}} |) r' \, \mathrm{d}r' \mathrm{d}\varphi',
\end{equation}
where  $\varphi'_{min}$ and $\varphi'_{max}$ correspond to the sectors of the sensor and $r'_{n}$~to the radius of the sensor pad $n$. 
The integration over $\varphi'$ comprises the sectors L1 and R1 (see Figure~\ref{fig_si_sensor}) which corresponds to about~40~mm. Since the transverse size of the beam is~$\sigma_{x,y}\approx$~4.2~mm and the expected effective Moli\`ere radius is around~10~mm, the two sectors safely cover one effective Moli\`ere radius of the shower.    

The trial function used to describe the average transverse energy profile of the shower is a Gaussian for the core, dominated by the high energy component of the shower, and a form inspired by the Grindhammer-Peters
parameterisation~\cite{Grindhammer,Grindhammer1} to account for the tails originating from the low energy photon halo,
\begin{equation} \label{eq_MR_FrFinel}
F_{E} ( r ) =  A_C e^{-(\frac{r}{R_C})^2} +  A_T \frac{2r^{\alpha}R_T^2}{ (r^2 + R_T^2 )^2 } \ ,
\end{equation}
where $A_C$, $R_C$, $A_T$, $R_T$ and $\alpha$ are parameters to be determined by fitting the function to the measured distribution. 

As can be seen from eqn.(\ref{eq_pad_energy_int_rphi}), the energy $E_{n}$ deposited in the pad number~$n$ depends on the shower position $\vec{R'_{0}}$ and pad position $\vec{r'_{n}}$. Since the beam transverse size is significantly smaller than the radius ${R'_{0}}$, the calculation of $E_{n}$ is done for a value of ${R'_{0}}$ which corresponds to the position of the maximum in the beam profile. This maximum is observed in a pad with $n$~=~45 and $r'_{n}$~=~161~mm. 

\subsection{The effective Moli\`ere radius determination at 5~GeV}

\begin{figure}[h!]
  \begin{minipage}[t]{0.45\textwidth}
      \includegraphics[width=\textwidth]{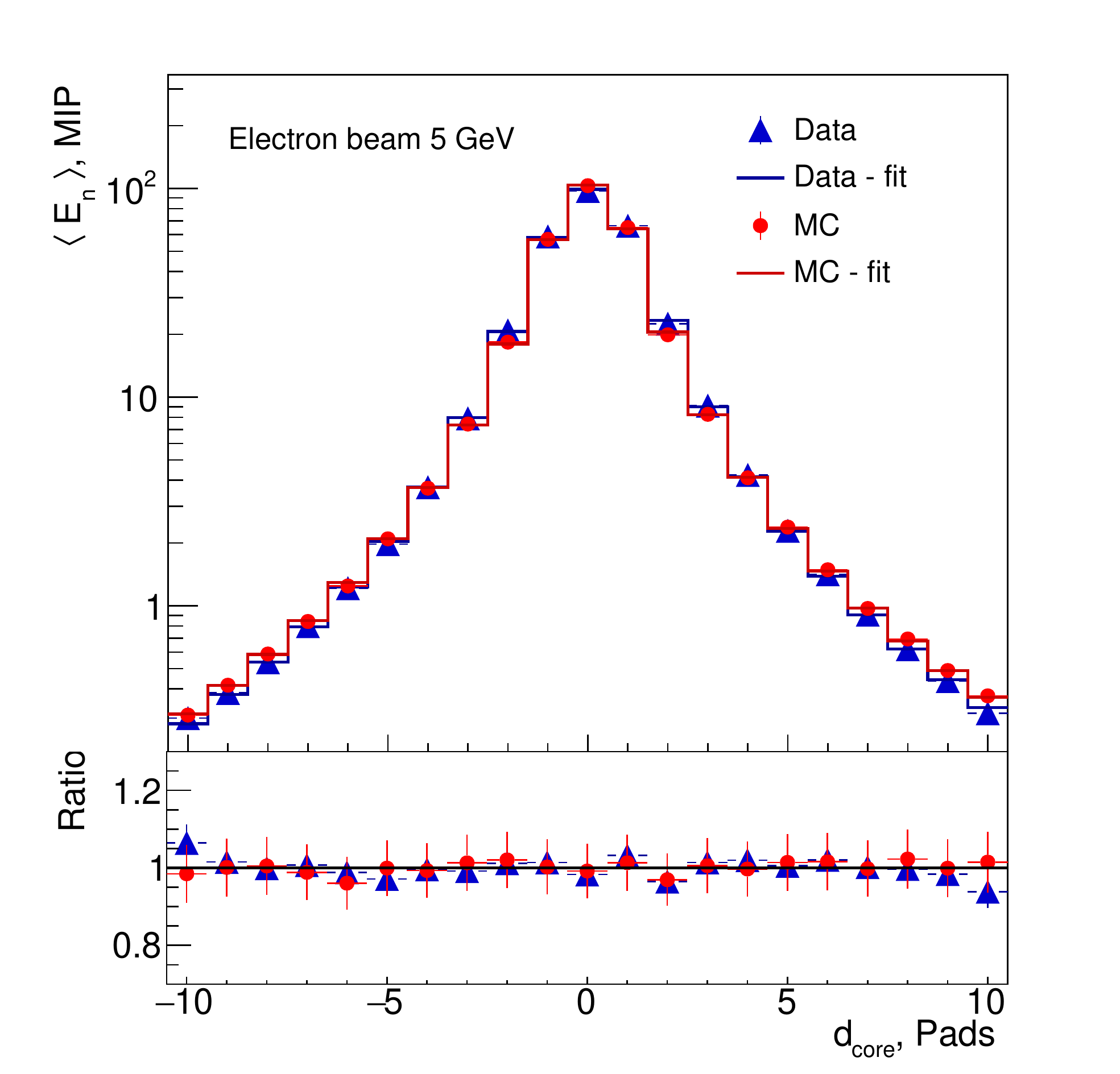}
    \caption{The average transverse shower profile, $\langle E^{det}_{n} \rangle$, as a function of the distance from the core, $d_{core}$, in units of the pad dimension (1.8~mm), for data (blue triangles) and MC simulation (red circles). The histograms are the results of fits to data 
    and Monte Carlo simulation to the function in eqn.(\ref{eq_MR_FrFinel}). The distributions are obtained with a 5 GeV electron beam. The lower part of the figure shows the ratio of the distributions to the fitted function, for the data (blue) and simulation (red).   }
    \label{MR_5GeV}
  \end{minipage}\hfill
  \hspace{0.025\textwidth}
  \begin{minipage}[t]{0.45\textwidth}
      \includegraphics[width=\textwidth]{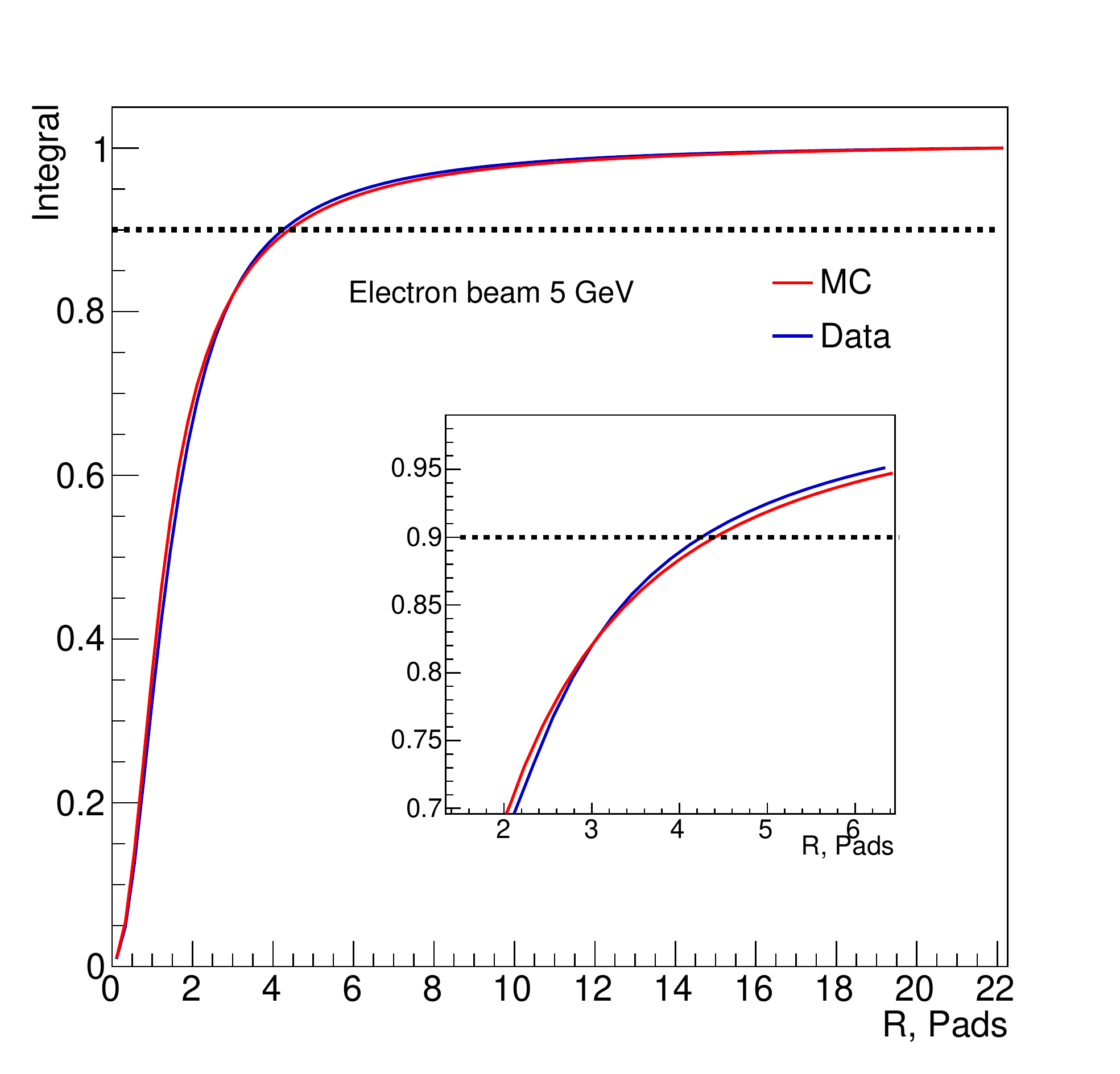}
    \caption{The ratio of the integrals in eqn~(\ref{eq_MR_1}) using $F_E(r)$ obtained from the fit, as a function of the radius $R$ in units of the pad dimension (1.8~mm), for data (blue) and MC (red), for a 5 GeV electron beam. The insert shows an expanded view of the region $2 < R < 6$ pads.} 
    \label{fig_integral}
  \end{minipage}
\end{figure}

The average profile of the electromagnetic transverse shower is determined by summing over all detector layers,  
\begin{equation} \label{eq_E_pad}
\langle E^{det}_{n} \rangle = \sum\limits_{l} \langle E^{det}_{nl} \rangle \ .
\end{equation}
The measured averaged transverse energy values, $\langle E^{det}_{n} \rangle$, were fitted to the function in eqn.(\ref{eq_MR_FrFinel}). Results for data and Monte Carlo simulation for electrons of 5 GeV energy, are shown in Figure~\ref{MR_5GeV}, where one sees the dependence of $\langle E^{det}_{n} \rangle$ on the distance from the shower core, $d_{core}$. The simulation agrees well with the data.
 The fitted function reproduces the experimental and the simulated transverse shower profile with an accuracy better than 5\%. Figure~\ref{fig_integral} shows the right part of eqn.~(\ref{eq_MR_1}) as a function of the radial integration limit~${R}$ for data and simulation with the horizontal line demonstrating a graphical solution for the effective Moli\`ere radius. The result is (8.1 $\pm$ 0.1 (stat) $\pm$ 0.3 (syst))~mm, a value well reproduced by the MC simulation (8.4 $\pm$ 0.1)~mm. The result obtained here is much smaller than 
 the one determined in the calorimeter prototype used during the 2014 test beam with larger gaps between the tungsten plates, which yielded (24.0 $\pm$ 1.6)~mm~\cite{LumiCal_multilayer_tb2014_epjc}.

\subsection{Energy dependence of the effective Moli\`ere radius}

\begin{figure}[h!]
  \begin{minipage}[t]{0.45\textwidth}
    \includegraphics[width=\textwidth]{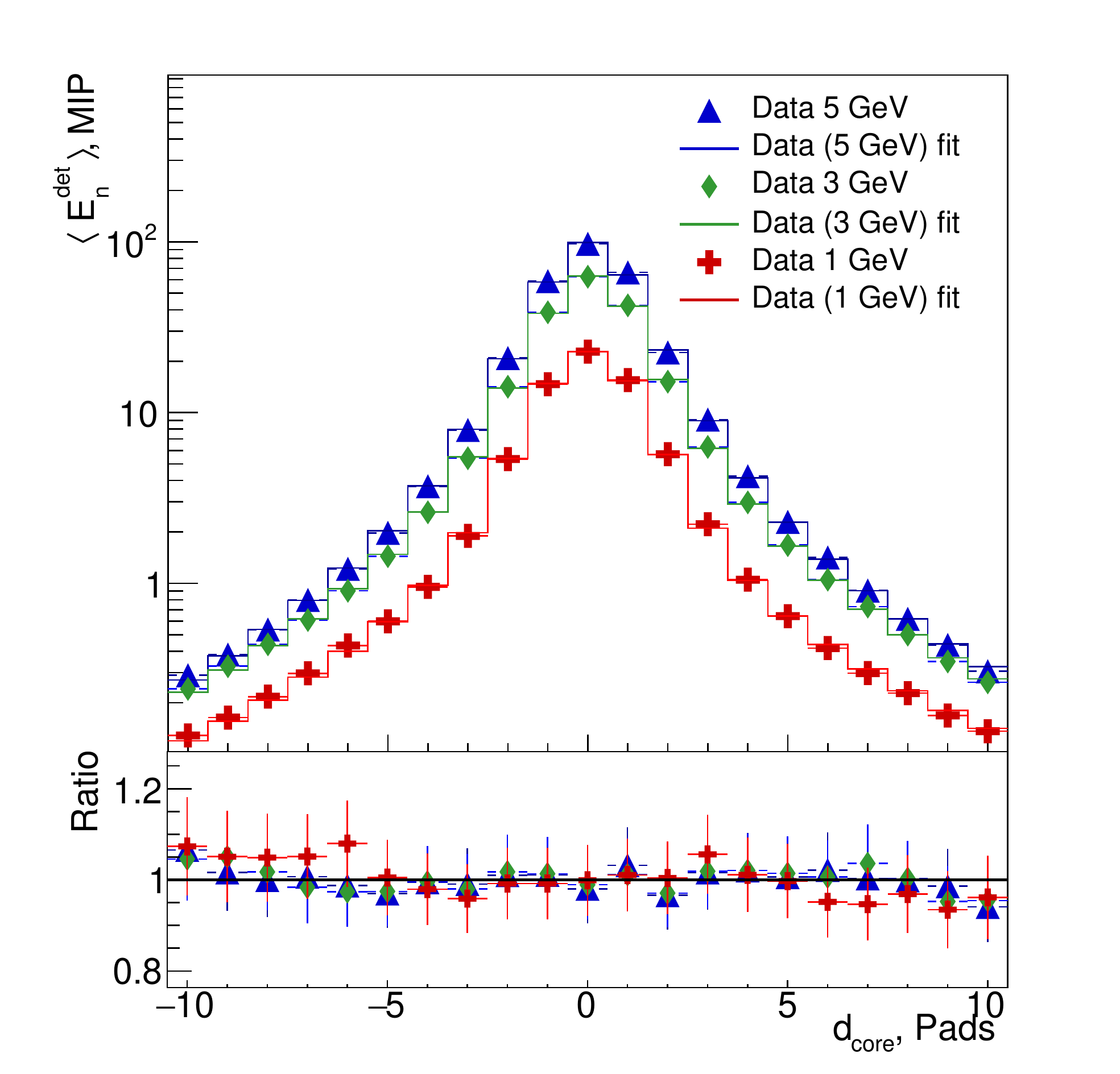}
    \caption{ Average transverse shower profiles for 1, 3 and 5 GeV electrons in data and the ratios between data and the fitted functions.}
    \label{fig_MR_Data135}
  \end{minipage}\hfill
  \begin{minipage}[t]{0.53\textwidth}
    \includegraphics[width=1.0\textwidth]{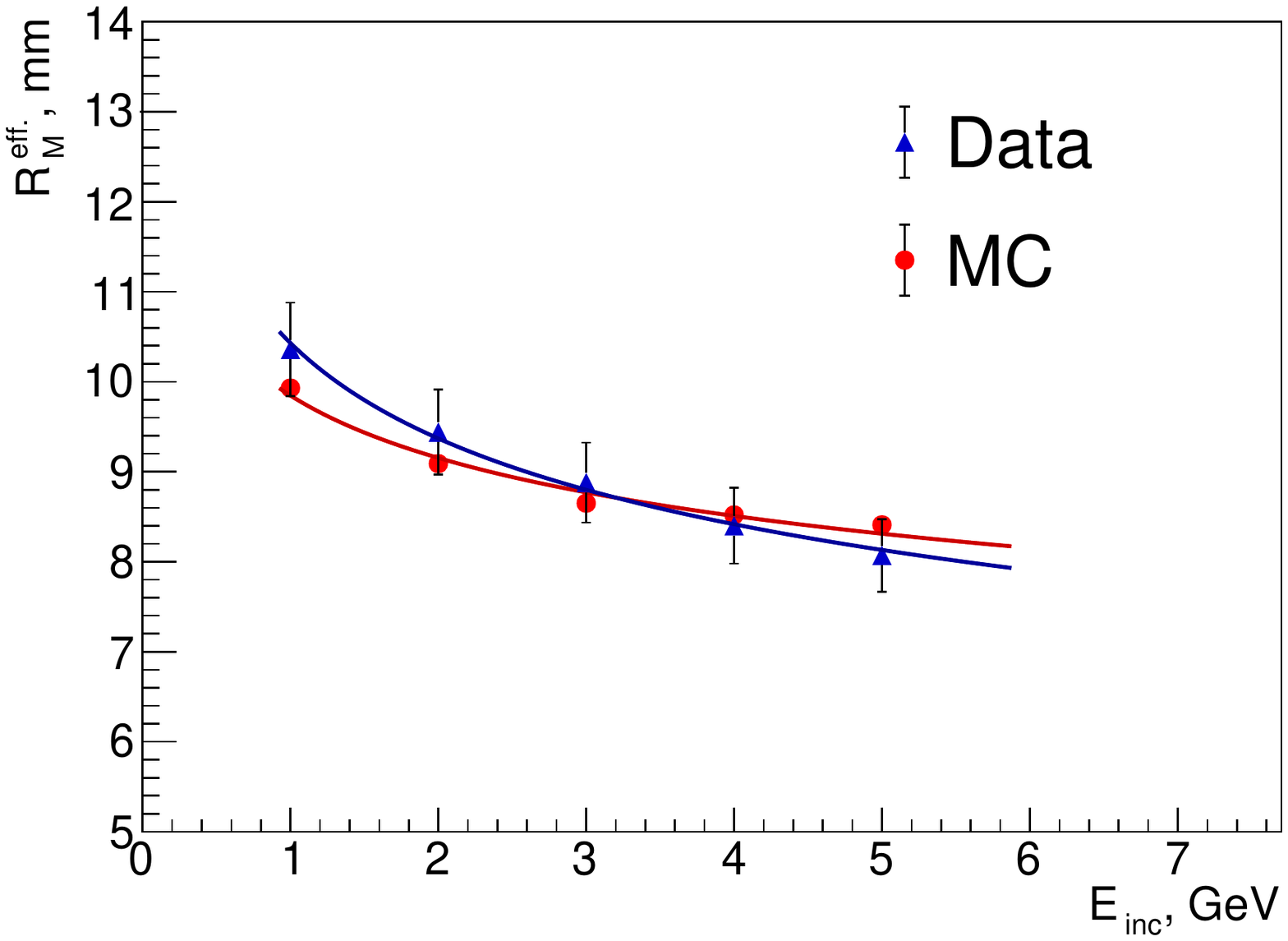}
    \caption{The effective Moli\`ere radius as a function of the electron energy for data (blue) and simulation (red).}
    \label{fig_MR_Edep}
  \end{minipage}\hfill
\end{figure}

The main analysis was performed for data taken at 5~GeV beam energy. In addition, data were taken for energies between 1 and 5~GeV. 
For the study of the energy dependence, about 50,000 events were used for each energy, and the measurement of the effective
Moli\`ere radius was carried out as for the 5~GeV sample. An example of the average transverse shower profiles at 1, 3 and 5~GeV beam energy  
is shown in Figure~\ref{fig_MR_Data135}. The average deposited energies are lower at lower beam energies, and the distributions are wider, 
resulting in a larger value of the effective Moli\`ere radius. The data are again well described by the results of simulations.

The effective Moli\`ere radius as a function of the incoming electron energy , $E_{inc}$, in the range of~1~-~5~GeV is shown in Figure~\ref{fig_MR_Edep}. It decreases with the electron energy as $E_{inc}^{(-0.15\pm 0.04)}$. The fit to the simulation yields an exponent (-0.11$\pm$0.01), in agreement with the data.

In order to investigate the observed energy dependence of the effective Moli\`ere radius, a simulation of an "infinite" calorimeter was performed. In practice the simulated calorimeter consisted of 30 planes with transverse size of 40 $\times$ 40~cm$^2$. Absorbers, detector layers and gaps had the same composition and thickness as the ones of the tested prototype.  

\begin{figure}[h!]
  \begin{minipage}[t]{0.45\textwidth}
      \includegraphics[width=\textwidth]{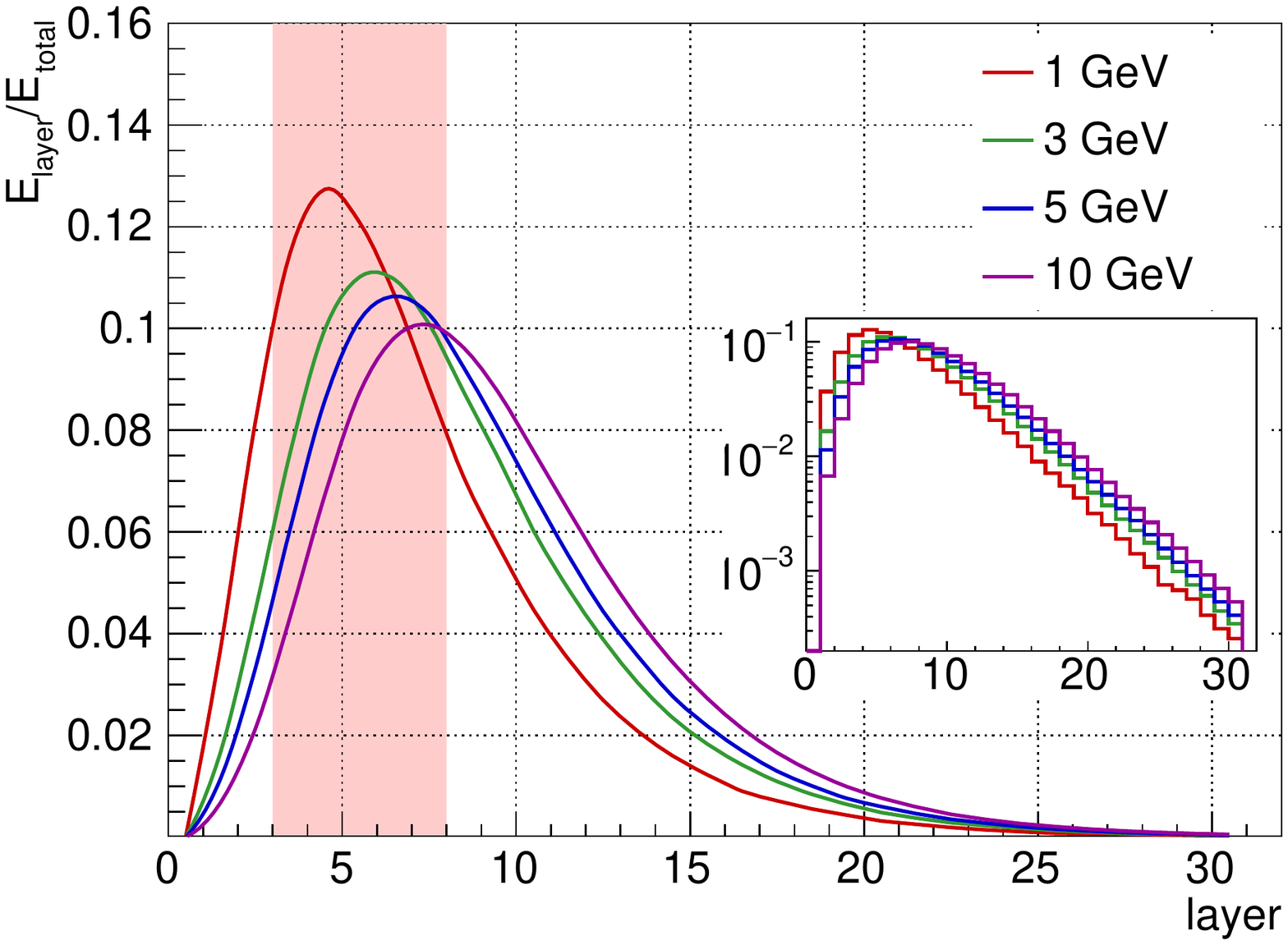}
    \caption{Normalised longitudinal shower profile for different electron energies obtained in simulation. Highlighted region corresponds to sensitive layers installed in the calorimeter. The inset shows the same profiles in a logarithmic scale.}
    \label{fig_l_shower_MC_1_3_5_10_GeV}
  \end{minipage}\hfill
  \hspace{0.09\textwidth}
  \begin{minipage}[t]{0.45\textwidth}
      \includegraphics[width=\textwidth]{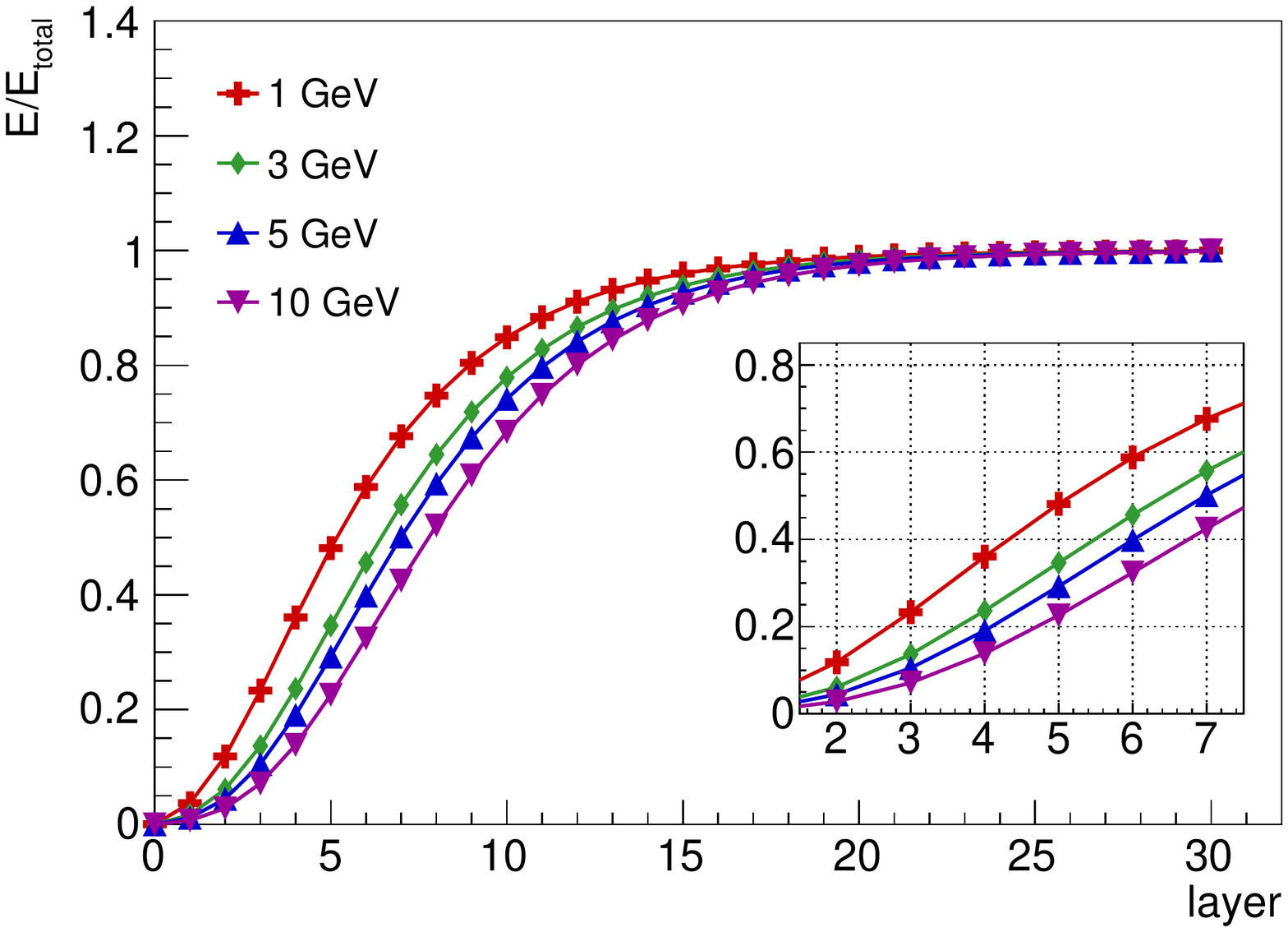}
    \caption{Cumulative distribution of the fraction of energy deposited in the detector layers as a function of the number of layers for different electron beam energies. The insert shows an expanded view of the region for planes 2 to 7.}
    \label{fig_e_lfraction_MC_1_3_5_10_GeV}
  \end{minipage}
\end{figure}

Figure~\ref{fig_l_shower_MC_1_3_5_10_GeV} shows the normalised average longitudinal profile of the energy deposited in the detector layers for incident electrons of different energy. The depth of the calorimeter is sufficient to contain most of the shower even for 10~GeV electrons in which case the fraction of the energy deposited in the last sensor layer is below 0.3\%, as can be seen in the insert in Figure~\ref{fig_l_shower_MC_1_3_5_10_GeV}. 
The detector layers from 3 to 7, as installed in the prototype   (shaded area) probe different regions of the longitudinal shower profile for different energies. For 1~GeV electrons, the shower is measured almost symmetrically around its maximum, while for 5~GeV electrons the layers~3~--~7 cover mostly the left side from the maximum. Hence, the fraction of the energy recorded in these layers depends on the beam energy. 
In Figure~\ref{fig_e_lfraction_MC_1_3_5_10_GeV} the cumulative distribution of the fraction of the deposited energy is shown as a function of the number of layers. In layers 3 to 7, the fractions for 1~GeV, 3~GeV and 5~GeV electrons are~56\%, 50\% and~46\%, respectively. 
This difference explains a small deviation from linearity in the observed prototype response as was shown in Figure~\ref{Avarage_E_dep_total_1_5gev} with red line and squares. Those measurements corrected to represent equal fractions of beam energies are shown with blue triangles and they are in good agreement with linear fit.

\begin{figure}[h!]
  \begin{minipage}[t]{0.45\textwidth}
    \includegraphics[width=\textwidth]{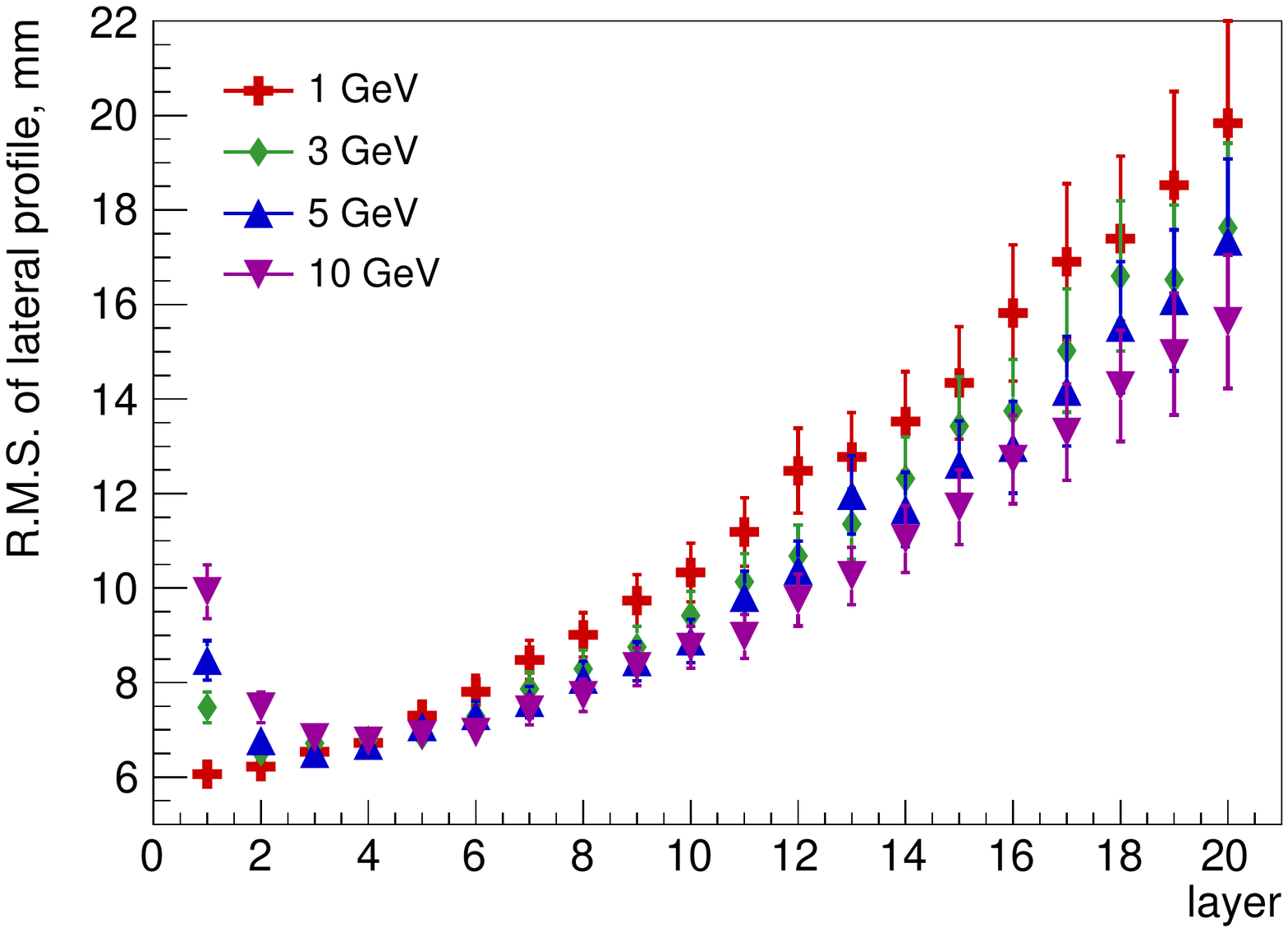}
    \caption{RMS of the lateral deposited energy distributions in detector layers obtained in simulation, for different electron beam     energies.}
    \label{fig_RMS_MC_1_3_5_10_GeV}
  \end{minipage}
  \hspace{0.09\textwidth}
  \begin{minipage}[t]{0.45\textwidth}
      \includegraphics[width=\textwidth]{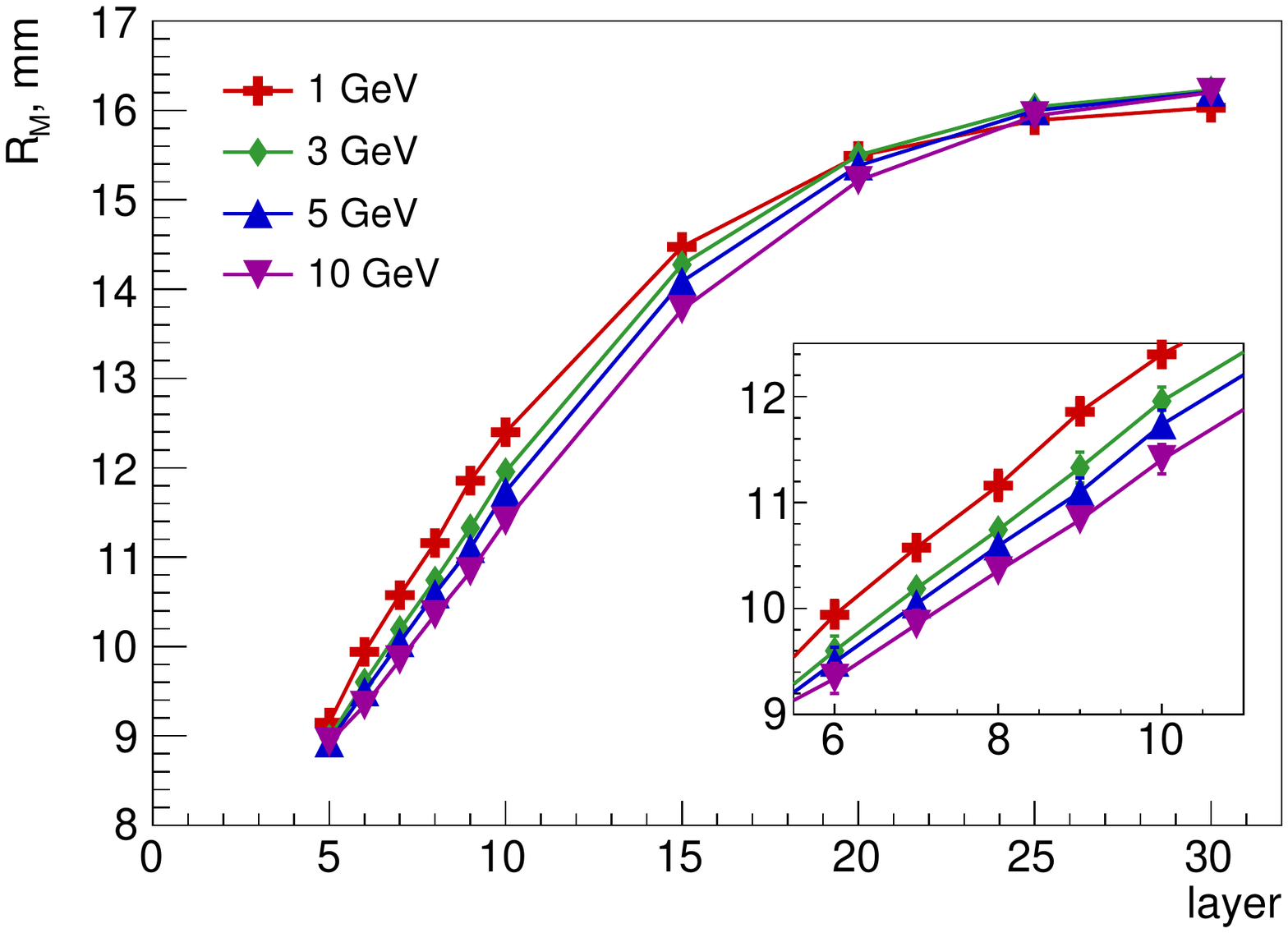}
    \caption{Moli\`ere radius, obtained in simulation, for different electron beam energies as a function of number of detector layers 
    in the calorimeter. The insert shows an expanded view of the plane region 6 - 10.}
    \label{fig_MR_MC_1_3_5_10_GeV}
  \end{minipage}\hfill
\end{figure}

The measurement of the shower in fixed detector layer positions for different longitudinal shower profiles also influences  the observed transverse shower size. As can be seen from Figure~\ref{fig_RMS_MC_1_3_5_10_GeV}, the RMS of the lateral projection of the deposited energy in each detector layer is expected to increase as a function of the sensor layer number, with a steeper slope for lower electron energies. The small increase of the RMS observed in the first and second layers are explained by the back-scattering of shower particles. According to the results in Figure~\ref{fig_RMS_MC_1_3_5_10_GeV} it is expected that the effective Moli\`ere radius decreases with increasing beam energy for the beam test geometry. When the fraction of the sampled shower energy approaches unity for different electron energies, the Moli\`ere radii converge to the same value. This can be seen in Figure~\ref{fig_MR_MC_1_3_5_10_GeV}, where the calculated Moli\`ere radius is shown as a function of the number of detector layers included in the calculation. Thus, the observed dependence of the effective  Moli\`ere radius in the prototype on the incident electron energy, as presented in Figure~\ref{fig_MR_Edep}, is due to the limited number of detector layers installed near the shower maximum. The slightly higher values of the effective Moli\`ere radius observed in the simulated calorimeter originate from the fact that in the simulation the transverse size of the calorimeter was much larger than that of the prototype. The difference is well reproduced by the simulation.

\subsection{Uncertainties}

The study of the systematic uncertainty of the measured average energy deposition in the transverse direction~$\langle E^{det}_{nl} \rangle$ includes the following contributions: 
\begin{itemize}
\item uncertainty of the measured efficiency of the signal identification; 
\item uncertainty of the particle impact position measurement and misalignment of detector planes;
\item uncertainty due to bad channels;
\item noise uncertainty;
\item calibration uncertainty.
\end{itemize} 

The uncertainty due to the efficiency of the signal reconstruction is evaluated by changing the efficiency according to high and low edges of the shaded area in Figure~\ref{fig_apv_efficiency}. The result for the effective Moli\`ere radius changes by $\pm$0.16~mm. 

The misalignment of the detector planes is estimated using occupancy plots for each layer. It is accounted for in the geometry of the simulation. The effect of misalignment on the effective Moli\`ere radius comes from the sum in eqn.(\ref{eq_E_pad}) where the radial pad index~$n$ denotes pads in different layers which are assumed to be aligned in the longitudinal direction. Due to misalignment, the average lateral deposited energy~$\langle E^{det}_{n} \rangle$ for a given distance from the shower core, determined by the index~$n$, gets contribution from pads which are at different distances from the shower core. A similar effect arises from the uncertainty of the particle impact position. This uncertainty is
estimated by calculating the effective Moli\`ere radius from simulations with perfectly aligned sensors and sensors displaced within the estimated misalignment. The change of the effective Moli\`ere radius is found to be 0.08~mm. 

The influence of the bad channels, which are included into simulation, leads to a change of the effective Moli\`ere radius by 0.14~mm compared to the simulation where all channels work properly.  

The effect due to the usage of one single radius~${R'_{0}}$ in equation~(\ref{eq_pad_energy_int_rphi}) for the calculation of~$E_{n}$ is estimated by selecting a narrow range of the particles impact position around the sensor pad with the radial index~$n=45$. The relative changes of the effective Moli\`ere radius is within~0.13~mm.  

The contribution of the measured noise uncertainty was studied in the simulation and found to be significantly below 1\%.

A relative calibration uncertainty of~5\% for each APV25 front-end board is assigned to each value~$\langle E^{det}_{nl} \rangle$ in eqn.~(\ref{eq_E_pad}) and summed in quadrature to determine the uncertainty of~$\langle E^{det}_{n} \rangle$. The calibration uncertainty is combined with the statistical one and used to produce 1000 transverse shower profiles where each~$\langle E^{det}_{n} \rangle$ is randomly generated using a Gaussian distribution function with a mean value corresponding to the measured~$\langle E^{det}_{n} \rangle$ and a $\sigma$ determined by the uncertainty. For each shower, the effective Moli\`ere radius is calculated and the RMS of their distribution is considered as a contribution to the statistical uncertainty of the effective Moli\`ere radius measurement.

The contributions to the systematic uncertainty are considered to be independent. The total systematic uncertainty on the Moli\`ere radius measurement is obtained by adding all the contributions in quadrature.

\section{Summary and conclusions}
\label{Summary_section}
New sub-millimeter thickness detector layers for the luminosity calorimeter LumiCal have been designed and produced. 
Silicon sensors are read out using Kapton fan-outs with copper traces connected via wire bonding or TAB to the sensor pads. 
The eight assembled detector layers were installed in the 1~mm gap between the tungsten absorber plates and successfully operated  during the 2016 beam-test campaign. 
Measurements of the shower position and the longitudinal and transverse shower shape are presented and compared to Monte Carlo simulations.
The effective Moli\`ere radius of this compact calorimeter prototype was determined at 5~GeV to be (8.1 $\pm$ 0.1 (stat) $\pm$ 0.3 (syst))~mm, a value well reproduced by the MC simulation (8.4 $\pm$ 0.1)~mm. Its energy dependence in the range 1 - 5~GeV was also studied. The observed slight decrease proportional to $E_{inc}^{(-0.15\pm 0.04)}$, can be explained by the limited number of detector planes used to probe the electromagnetic shower.

These results demonstrate the feasibility of constructing a compact calorimeter consistent with the conceptual design, which is optimised for a high precision luminosity measurement in future e$^{+}$e$^{-}$ collider experiments.
%
\section*{Acknowledgements}
This study was partly supported by the Israel Science Foundation (ISF), Israel German Foundation (GIF), the I-CORE program of the Israel Planning and Budgeting   Committee, Israel Academy of Sciences and Humanities, by the National Commission for Scientific and Technological Research (CONICYT - Chile) under grant FONDECYT 1170345, by the Polish Ministry of Science and Higher Education under contract nrs 3585/H2020/2016/2 and 3501/H2020/2016/2, the Rumanian UEFISCDI agency under contracts PN-II-PT-PCCA-2013-4-0967 and PN-II-ID-PCE-2011-3-0978, by the Ministry of Education, Science and Technological Development of the Republic of Serbia within the project Ol171012, by the United States Department of Energy,
grant DE-SC0010107, and by the European Union Horizon 2020 Research and Innovation programme under Grant Agreement no.654168 (AIDA-2020). The measurements leading to these results have been performed at the Test Beam Facility at DESY Hamburg (Germany), a member of the Helmholtz Association (HGF)

\end{document}